# Entropic Attraction of Heavy Spheres in a Harmonically Driven Bath of Poppy Seeds

Juleon M. Schins

Guest scientist at the "Departamento de Física y Matemática Aplicada,

Facultad de Ciencias, Universidad de Navarra", Pamplona, Spain

***Abstract***

*We develop the simplest possible theory that gives reason of the recent experimental observations that two heavy spheres, immersed into a monolayer bath of poppy seeds, attract one another when shaken harmonically in a single horizontal direction. Their attraction is so strong that the two spheres remain bound during hundreds of driving periods. The paper consists of three independently readable Chapters, with only a few inter-chapter references. The first Chapter concerns itself with the the motion of a roller amidst a horizontally shaken sea of poppy seeds, under the Ansatz of equal phase for rotational and translational velocities. The second Chapter details how this predicts the observed longitudinal diffusion of a single sphere in a bath of poppy seeds. The third Chapter shows how to retrieve all relevant physical parameters from experiment: viscosity of the seed bath, friction of the heavy spheres with the harmonically moving substrate, equilibrium rates for dissociation of bound pairs or binding of lone pairs, and the full Gibbs potential surface as a function of experimentally accessible parameters.*

PACS numbers : 45.70.Mg, 05.40.-a

# Chapter I:

# Concentration Dependence of a Sphere's Longitudinal Diffusion

## 1.1. Introduction

Monolayer Binary Segregation is the physical phenomenon that two arbitrary kinds of monodisperse particles (like heavy phosphor-bronze spheres and light poppy seeds) may cluster when shaken [1–3]; depending on the system parameters, this clustering is gas-like, liquid-like, or solid-like [ref PRE 74, Mullin]. Its mechanism depends on the friction coefficients, on particle size, density, and shape, and on the shaking motion of the underlying tray. Up to date, two principal mechanisms of segregation are (i) the "Brazil nut effect in vertical shaking" [4] or avalanches [5–7], where particles gather together at a particular location, and (ii) a physical attraction between similar particles without any preferred spatial location [8,9]. In the latter case, considerable debate has arisen over the nature of the forces that lead to the aggregation of similar particles. One possibility is an excluded volume effect, but this effect does not reach beyond one particle diameter, whence it fails to explain the experimental data in samples of dilute phosphor-bronze spheres [10,11]. Examples of long-range interactions result from pressure, density, or velocity fluctuations [12, 13, 14] in the region between clustering particles. A Casimir effect occurs on very long range, and can be both repulsive and attractive depending on the system parameters [15, 16], though it is too weak to explain the time scales of the formation of pairs of phosphor bronze spheres (still "gas phase"). Much experimental work has recently appeared on heavy spheres immersed in a harmonically shaken horizontal monolayer of dry poppy seeds [17–24].

This contribution has three Chapters. In Chapter I, we calculate the response of, separately, a monolayer seed bath and a single heavy sphere on an empty tray, to harmonic driving. In Chapter II, we calculate the longitudinal diffusion of a single heavy sphere in a shaken monolayer bath of poppy seeds. In Chapter III, we estimate the physical parameters determining attractive force between two heavy spheres in the same shaken monolayer bath.

## 1.2. Ancestors of the Stick-Free Attractor

This paper uses the same conventions and calculation procedures as Ref. 30, which focused on the response of a heavy slider immersed in a shaken monolayer seed bath. Here too, we start by calculating the "stick-free attractor" (that is, the idealized response assuming no sticking nor stalling), not of a slider, but of a roller, like a sphere or a rod. Conservation of momentum requires

[2.1] $$F_{angular} + F_{inertia} = F_{friction} + F_{ext}$$

with

[2.2] $$\begin{Bmatrix} F_{friction} = -m\mu_{dyn} g\,\mathrm{sgn}[v^{[rel]} + R\omega] \\ F_{inertia} = m\dot{v}^{[lab]} \\ F_{angular} = I\dot{\omega}/R = mJR\dot{\omega} \end{Bmatrix}$$

with $v^{[lab]} \equiv v^{[rel]} + u$ the laboratory and driver-referenced velocities of the roller, $u$ the lab-frame velocity of the moving substrate (henceforth called "driver"), $\omega$ the angular velocity of the object, $J \equiv \frac{I}{mR^2}$ the reduced angular inertia, $R$ the roller's diameter, and $m$ its mass; with $\mu_{stat}$ and $\mu_{dyn}$ the static and dynamic friction coefficients, respectively; and $g$ the earth's gravitational constant.

Eqs 2.1 and 2.2 state that

(i) the force of kinetic friction, e.g., of a slider, acts like the break on motion, and tends to reduce the slider's relative velocity to zero, as it is always directed opposite to the direction of the velocity;

(ii) according to Newton, an object undergoing a force acting on its center of mass, only changes its lab-frame linear velocity;

(iii) again according to Newton, a 2D-circle undergoing an in-plane force perpendicular to the radius, changes its rotational velocity;

(iv) the non-slipping condition is $v^{[rel]} + R\omega = 0$.

Henceforth, we shall omit the [lab]-superscript on the laboratory velocity, and leave the [rel]-superscript only for the tray-referenced velocity. For a roller, be it a sphere or a cylinder, on an immobile substrate ($u = 0$), and in the absence of any external forces, the conservation of momentum law reduces to

[2.3] $$JR\dot{\omega} + \dot{v} + \mu_{dyn} g \,\mathrm{sgn}[v + R\omega] = 0$$

The simplest solution has constant accelerations $\dot{v}_0$ and $\dot{\omega}_0$:

[2.4] $$JR\dot{\omega}_0 + \dot{v}_0 + \mu_{dyn} g \,\mathrm{sgn}[v + R\omega] = 0$$

For this equation of motion, one may define the "free-flight time" $t_{ff}$ as the time needed to overcome slipping:

[2.5] $$v_0 + t_{ff}\dot{v}_0 + R(\omega_0 + t_{ff}\dot{\omega}_0) = 0$$

whence

[2.6] $$t_{ff}(v_0, \omega_0, \dot{v}_0, \dot{\omega}_0) = -\frac{v_0 + R\omega_0}{\dot{v}_0 + R\dot{\omega}_0}$$

The solutions seem to be independent of the roller's angular momentum:

[2.7] $$\begin{Bmatrix} v(t) = (v_0 + t\dot{v}_0)\theta(t_{ff} - t) + (v_0 + t_{ff}\dot{v}_0)\theta(t - t_{ff}) \\ \omega(t) = (\omega_0 + t\dot{\omega}_0)\theta(t_{ff} - t) + (\omega_0 + t_{ff}\dot{\omega}_0)\theta(t - t_{ff}) \end{Bmatrix}$$

However, the initial condition quartet $\{v_0, \omega_0, \dot{v}_0, \dot{\omega}_0\}$ must satisfy

[2.8] $$JR\dot{\omega}_0 + \dot{v}_0 + \mu_{dyn} g \,\mathrm{sgn}[v_0 + R\omega_0] = 0$$

As long as $t < t_{ff}$, slipping occurs; this implies changing velocities, and conversion of motional energy into heat. On the other hand, for $t > t_{ff}$, an equilibrium state is reached in which the roller rolls without friction, i.e., satisfying the condition that the velocity of the roller is equal to that of the substrate at the contact point:

[2.9] $$V + R\Omega = 0$$

Where we used the definitions

[2.10] $$\begin{Bmatrix} V \equiv v(t_{ff}) \\ \Omega \equiv \omega(t_{ff}) \end{Bmatrix}$$

From Eq. 2.5 it follows that the quartet of initial conditions $\{v_0, \omega_0, \dot{v}_0, \dot{\omega}_0\}$ has the same free-flight time as the quartet $\{v_0 + V_1, \omega_0 + \Omega_1, \dot{v}_0, \dot{\omega}_0\}$, provided $V_1 + R\Omega_1 = 0$. The only thing changing in the solution are the velocity offsets. Hence, all solutions to Eq. 2.4 can be divided into subfamilies of solutions with $R\Omega = V = 0$, and related to the ancestor like $\{v_0 + V_1, \omega_0 + \Omega_1, \dot{v}_0, \dot{\omega}_0\}$. The ancestors have the property

[2.11] $$t_{ff} = -\frac{v_0}{\dot{v}_0} = -\frac{\omega_0}{\dot{\omega}_0} \quad \Rightarrow \quad v_{kick}(t) = \theta(t) v_0 (1 - \frac{t}{t_{ff}}) \theta(t_{ff} - t); \quad \omega_{kick}(t) = \theta(t) \omega_0 (1 - \frac{t}{t_{ff}}) \theta(t_{ff} - t)$$

Hence, from the ancestor solutions one may reconstruct all other solutions by adding an arbitrary amount of velocities satisfying the no-slipping condition Eq. 2.7.

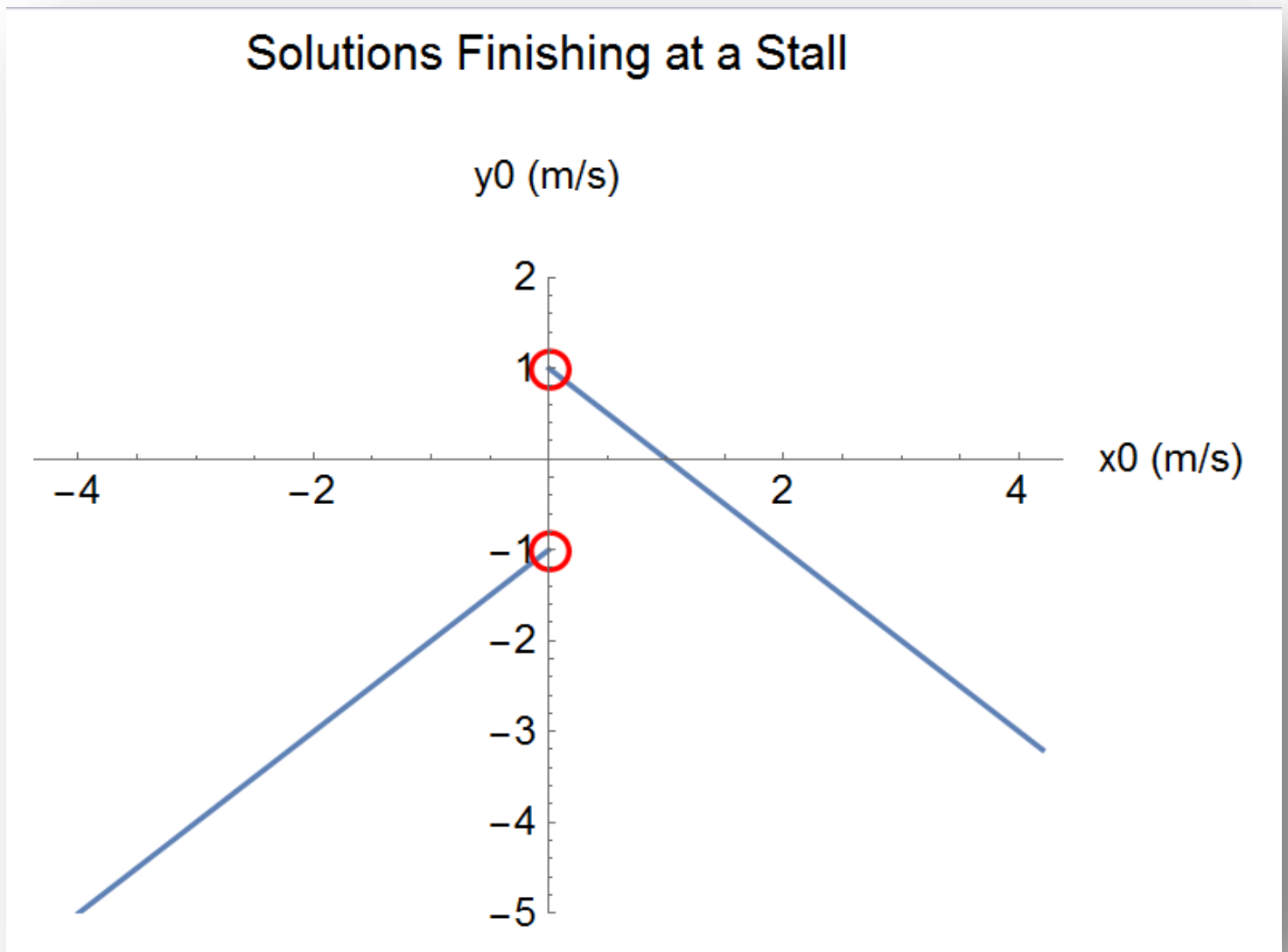

***Figure 2.1: Ancestor Solutions.*** *We assume* $2\mu_{dyn} g t_{ff} = 1 ms^{-1}$. *The red circles indicate the two points of the solution where* $v_0 + R\omega_0 = 0$. *The latter solutions are trivial, because they lack heat losses.*

From Eqs. 2.8 and 2.11 the ancestors also satisfy

[2.12] $JR\omega_0 + v_0 = \mu_{dyn} g t_{ff} \,\mathrm{sgn}[v_0 + R\omega_0]$

Consider the coordinate transformation

[2.13] $\begin{Bmatrix} x_0 \equiv (1+J)(v_0 + R\omega_0) \\ y_0 \equiv (1-J)(v_0 - R\omega_0) \end{Bmatrix} \quad \Rightarrow \quad y_0 = 2\mu_{dyn} g t_{ff} \,\mathrm{sgn}[x_0] - x_0 = \mathrm{sgn}[x_0](2\mu_{dyn} g t_{ff} - |x_0|)$

Fig. 2.1 illustrates solution Eq. 2.11 for $2\mu_{dyn} g t_{ff} \equiv 1 ms^{-1}$.

## 1.3. Convolution with the Ancestor

In this Section, we calculate the convolution of the ancestor solution presented in Section 2 with the driver's harmonic velocity. The common feature of these convolutions is that the rollers do not "roll off" when driven harmonically. The solutions are fine as long as Eqs. 2.11 and 2.12 are satisfied. That is, the initial condition quartets must be of the form

[3.1] $\{v_0, R\omega_0, \dot{v}_0, R\dot{\omega}_0\} = \{\mu_{dyn} g t_{ff} \,\mathrm{sgn}[v_0 + R\omega_0] - JR\omega_0, R\omega_0, -\frac{v_0}{t_{ff}}, -\frac{R\omega_0}{t_{ff}}\}$

The kernel is independent of those initial conditions:

[3.2] $K(t \,|\, t_{ff}) = \frac{2}{t_{ff}}(1 - \frac{t}{t_{ff}})\theta(t)\theta(t_{ff} - t) \quad \Rightarrow \quad \lim_{t_{ff} \downarrow 0} K(t \,|\, t_{ff}) = \delta(t)$

The harmonically driven stick-free velocities are

[3.3] $\begin{Bmatrix} R\omega^{[stick-free]}_{harm-ancestor}(t \,|\, t_{ff}) = v^{[stick-free]}_{harm-ancestor}(t \,|\, t_{ff}) = u_0 \int_0^{t_{ff}} ds\, K(s) \cos\omega_h (t-s) = \\ = 2\frac{u_0}{t_{ff}} \int_0^{t_{ff}} ds\, (1 - \frac{s}{t_{ff}}) \cos\omega_h (t-s) = u_0 A_h \cos(\omega_h t - \phi_h) \end{Bmatrix}$

with

[3.4] $$\begin{Bmatrix} A_c \equiv 1 - \cos\omega_h t_{ff} \\ A_s \equiv \omega_h t_{ff} - \sin\omega_h t_{ff} \end{Bmatrix} \quad \begin{Bmatrix} \tfrac{1}{2}(\omega_h t_{ff})^2 A_h \equiv \sqrt{A_c^2 + A_s^2} \\ \tan\phi_h \equiv \dfrac{A_s}{A_c} \end{Bmatrix}$$

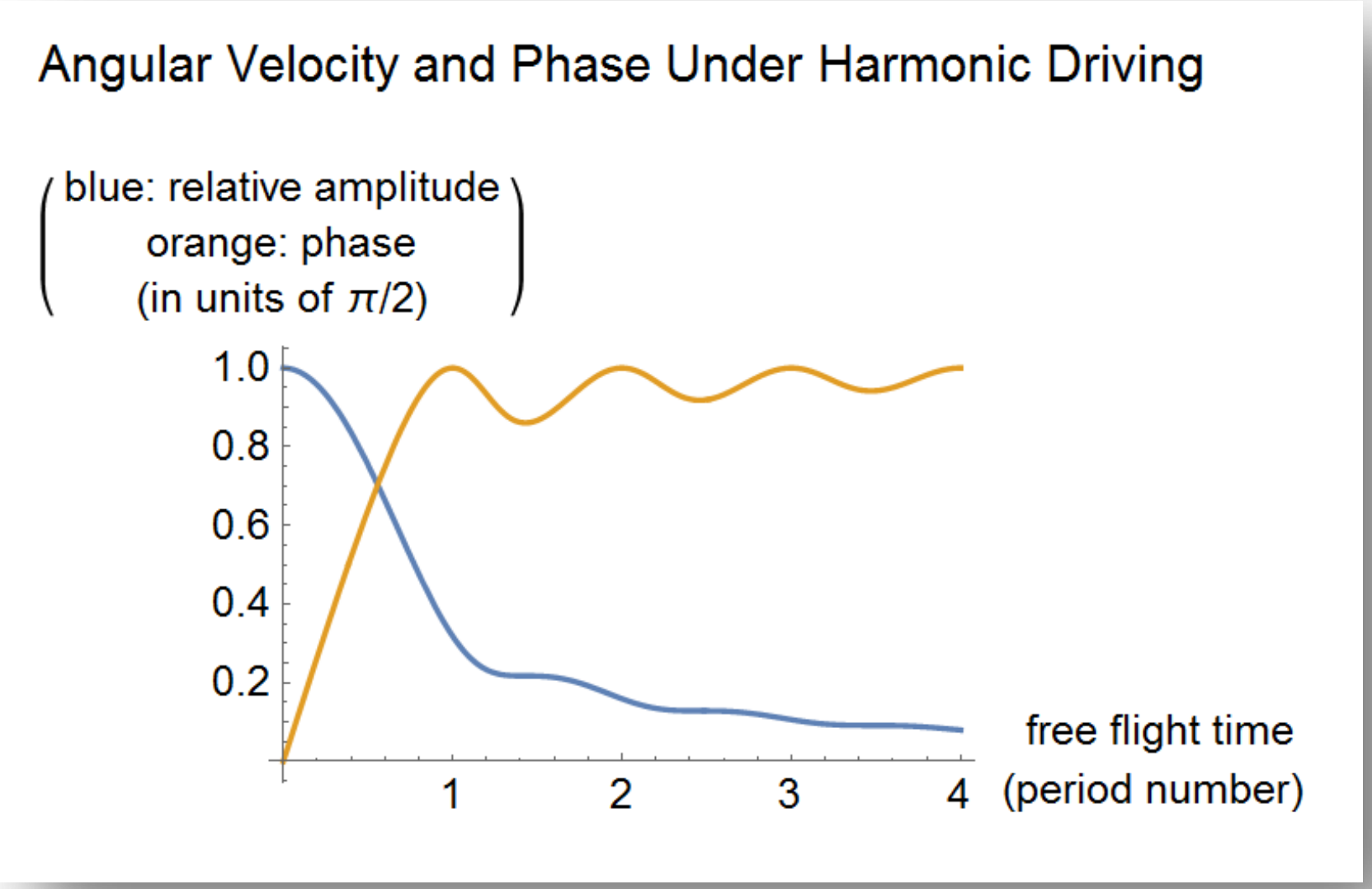

***Figure 3.1: Harmonic Responses of Both Velocities.*** *For infinitely long free-flight times the phase tends to π/2, and the relative amplitude to zero. The relative amplitude is equal to the driver's amplitude divided by the driver's. As for vanishing free flight time the kernel becomes a Dirac delta function, the slider is permanently stuck to the moving substrate for period number zero.*

Even the limits

[3.5] $$\begin{Bmatrix} \lim\limits_{\omega_h t_{ff} \to \infty} v^{[stick-free]}_{harm-ancestor}(t \,|\, t_{ff}) = 0 \\ \lim\limits_{\omega_h t_{ff} \to 0} v^{[stick-free]}_{harm-ancestor}(t \,|\, t_{ff}) = u_0 \end{Bmatrix}$$

are identical to the slider's case [30].

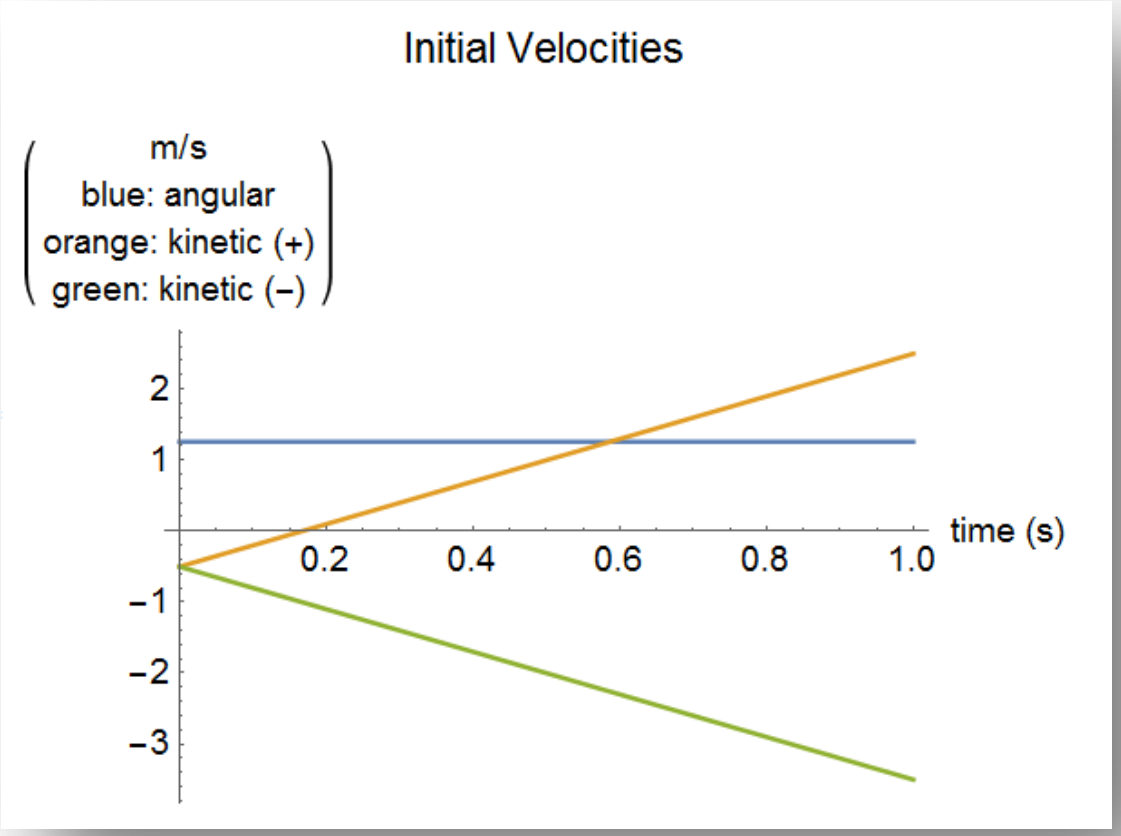

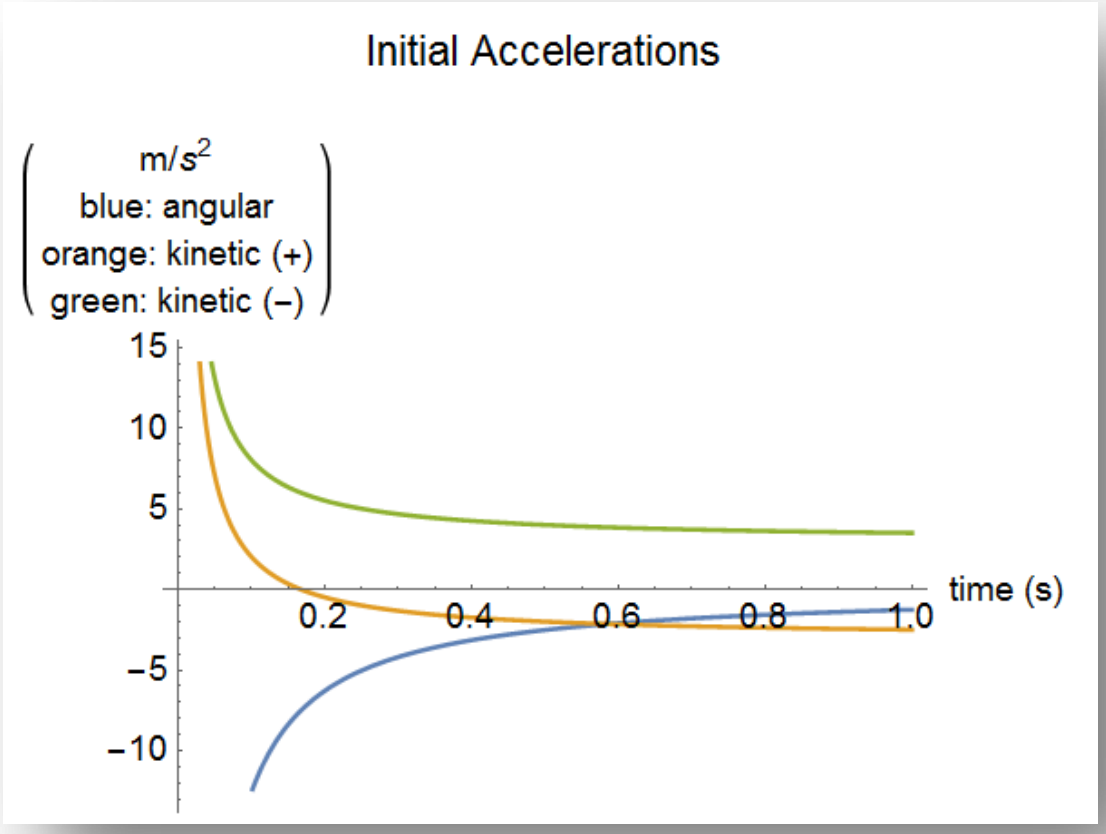

***Figure 3.2: Initial Velocities and Accelerations for either Sign of*** $v_0 + R\omega_0$***.*** *We used the values* $\mu_{dyn} g = 3ms^{-2}$ *and the reduced angular momentum* $J_{sphere} = 2/5$*. Eq. 3.1 underlies these plots.*

In Figs. 3.1 through 3.3 we plot the above deduced information. Note that only Fig 3.1 is important for the harmonic response. The other two figures relate initial velocities and accelerations satisfying the zero roll-off condition.

## 1.4. 1D-Momentum Conservation

Consequently, all harmonic responses that do not roll off with respect to a reference have the property that angular and kinetic velocities are identical in magnitude and phase. This means that the same Markovian model is applicable as in the case of a slider [30], although with a much less important weight for the static friction. In practice, spheres are are often embedded in a bath of seeds, which are about an order of magnitude lighter than the (phosphor-bronze) spheres. The seeds often

(i) have high friction coefficients due to their rugged shape;
(ii) are sliders rather than rollers (as are the spheres).

Suppose we know the response of either kind of particle in an otherwise empty tray. From the previous three Sections we know that the roller's response is harmonic even for large differences between static and dynamic friction. The question of interest is what the velocities of the single sphere and many seeds are, when mixed on a single tray. Assume the concentration and velocity amplitude of the seeds are high enough as to make stalling of seeds impossible, due to continuous bombardment of neighboring seeds. As far as the free flight time is concerned, we assume $t_r < t_R$, because the dynamic friction coefficients relate oppositely $\mu_r > \mu_R$, (see Eq. 2.12 for rollers, and $|v_0| = \mu_{dyn} g t_{ff}$ for sliders). Let us elaborate on the special case that $\omega_h t_r = 2\pi \times 0.3$ and $t_R = 4t_r$. According to Fig. 3.1 this implies $\phi_r = 0.19\pi$, $\phi_R = 0.47\pi$, $f_r^{[lab]} = 90\%$, and $f_R^{[lab]} = 23\%$. Note that the amplitudes are laboratory-frame referenced entities:

$$[4.2] \qquad f^{[lab]} \equiv \frac{v_{slider}^{[lab]}}{v_{d0}} \qquad \& \qquad f^{[rel]} \equiv 1 - f^{[lab]}$$

Hence, with respect to the moving substrate, the sphere has a much larger free swing ( $f_R^{[rel]} = 77\%$ ) than the poppy seeds ( $f_r^{[rel]} = 10\%$ ). This result, which is typical of most experiments performed in the field [1-24], shows that the spheres typically have to fight their way through a viscous sea of slower moving seeds. The seeds' 1D-longitudinal velocity distribution in 2D space is

$$[4.3] \qquad v_{rx}^2 = \frac{2}{m} k_B T_x$$

Momentum conservation ( $p_{com} = p_R + p_r$ ) dictates the inelastic longitudinal momentum transfer to the sphere in a 1D-seed-sphere collision

$$[4.4] \qquad \left\{ \begin{array}{l} p_{com-x} = m_R v_{Rx} + m_r v_{rx} \equiv (m_R + m_r) v_{com} \\ m_R (v_{Rx} - v_{com-x}) + m_r (v_{rx} - v_{com-x}) = 0 \end{array} \right\} \quad \Rightarrow \quad v_{com} \equiv \frac{m_R v_R + m_r v_r}{m_R + m_r}$$

whence

$$[4.5] \qquad v_{after} = 2 v_{com} - v_{before} \quad \Rightarrow \quad \left\{ \begin{array}{l} v_{Rx}^{[after]} = \dfrac{2 m_r v_{rx}^{[before]} + (m_R - m_r) v_{Rx}^{[before]}}{m_R + m_r} \\ v_{rx}^{[after]} = \dfrac{2 m_R v_{Rx}^{[before]} - (m_R - m_r) v_{rx}^{[before]}}{m_R + m_r} \end{array} \right\}$$

Upon adding a second dimension, Eq. 4.6 only holds for a head-on collision ($\Delta y = 0$), with $y$ the transverse coordinate. In case the collision is not head-on, but with the seed impinging $\Delta y < R$ off the sphere center's transverse coordinate, Eq. 4.5 transforms into

[4.6] $$v_{after}(\theta) = (2v_{com} - v_{before})\cos^2\theta$$

with $\theta$ the angle between the longitudinal or $x$-axis, and the line joining the sphere's center with the impact of a point-like seed particle onto the sphere's circumference. It relates to the transverse offset as

[4.7] $$R\sin\theta = \Delta y$$

The squared cosine in Eq. 4.6, averaged over the transverse direction, equals two thirds. Hence

[4.8] $$v_{Rx}^{[after]} = \frac{2}{3}\frac{2m_r v_{rx}^{[before]} + (m_R - m_r)v_{Rx}^{[before]}}{m_R + m_r}$$

This means that, for every collision with a 0 K seed bath ($v_{rx\&y}^{[bef]} = 0$), the sphere loses an average velocity

[4.9] $$\Delta v_{Rx} \equiv v_{Rx}^{[bef]} + v_{Rx}^{[aft]} = (\frac{m_R + m_r}{m_R + m_r} - \frac{2}{3}\frac{m_R - m_r}{m_R + m_r})v_{Rx}^{[bef]} = \frac{1}{3}\frac{m_R + 5m_r}{m_R + m_r}v_{Rx}^{[bef]}$$

## 1.5. Longitudinal Velocity of the Sphere

The equation of motion for the longitudinal velocities of the sphere is

[5.1] $$JR(\dot{\omega}_{Ry} + \gamma_{R\omega}\omega_{Ry}) + \dot{v}_{Rx}^{[rel]} + \gamma_{Rv}v_{Rx}^{[rel]} + \mu_{dyn}g\,\mathrm{sgn}[v_{Rx}^{[rel]} + R\omega_{Ry}] = \dot{u}$$

The subscript $x$ for the sphere's velocity, and $y$ for the radial frequency's axis of rotation, indicate the longitudinal nature. Eq. 5.1 further simplifies in case the two viscous friction coefficients are numerically equal ($\gamma_{R\omega y} = \gamma_{Rvx} = \gamma_R$):

[5.2] $$JR(\dot{\omega}_{Ry} + \gamma_R\omega_{Ry}) + \dot{v}_{Rx}^{[rel]} + \gamma_R v_{Rx}^{[rel]} + \mu_{dyn}g\,\mathrm{sgn}[v_{Rx}^{[rel]} + R\omega_{Ry}] = \dot{u}$$

The total longitudinal energy of motion of the sphere is

[5.3] $$E_{Rx} = E_{Rx-kin} + E_{Ry-rot} = \tfrac{1}{2}m(\overline{v_{Rx}^2} + JR^2\overline{\omega_{Ry}^2})$$

In the high temperature limit of the seed bath (valid for concentrations below 90%), energy equipartition for the sphere tends to equalize the two averages:

[5.4] $$\overline{v_{Rx}^2} = JR^2\overline{\omega_{Ry}^2}$$

Inspired by this high-temperature limit, we introduce the following *Ansatz*: At all times the two velocities have the same phase

[5.5] $$R\omega_{Ry} = -\beta v_{Rx}^{[rel]}$$

Clearly, for $\beta \neq 1$, there is a permanent power conversion into heat. Eq. 5.2 becomes

[5.6] $$\dot{v}_R^{[rel]} + \gamma_R v_R^{[rel]} + \xi\,\mathrm{sgn}[v_R^{[rel]}] = \frac{\dot{u}}{1 - J\beta}$$

with the introduction of $\beta$ and $\xi \equiv \dfrac{\mu_{dyn}g}{1 - J\beta}$, the former being a measure for how well energy equipartition is satisfied. Assume that Eq. 5.8 is a solution of Eq. 5.7.

[5.7] $$\dot{G} + \gamma_R G + \xi\theta(t)\theta(t_{ff} - t) = u_0\delta(t)$$

[5.8] $$\begin{cases} G(t) = G_0(e^{-\gamma_R t} - e^{-\gamma_R t_{ff}})\theta(t)\theta(t_{ff} - t) \\ \dot{G}(t) = G_0(1 - e^{-\gamma_R t_{ff}})\delta(t) - G_0\gamma_R e^{-\gamma_R t}\theta(t)\theta(t_{ff} - t) \end{cases}$$

Substitution of Eq. 5.8 into Eq. 5.7 yields

[5.9] $$(1 - e^{-\gamma_R t_{ff}})\delta(t) + [\gamma_R(e^{-\gamma_R t} - e^{-\gamma_R t_{ff}}) - \gamma_R e^{-\gamma_R t} + \frac{\xi}{G_0}]\theta(t)\theta(t_{ff} - t) = \frac{u_0}{G_0}\delta(t)$$

This is consistent only if

[5.10] $$\begin{cases} G_0\gamma_R e^{-\gamma_R t_{ff}} = \xi \\ G_0(1 - e^{-\gamma_R t_{ff}}) = u_0 \end{cases} \quad \Rightarrow \quad \begin{cases} e^{\gamma_R t_{ff}} = \frac{\gamma_R}{\xi}G_0 = 1 + \frac{\gamma_R u_0}{\xi} \\ G_0 = \frac{\xi}{\gamma_R} + u_0 \end{cases}$$

Eq. 5.7 transforms into Eq. 5.6 by a convolution

[5.11] $$\dot{u}(t) = \int_{-\infty}^{\infty} ds\, \frac{\dot{u}(t-s)}{u_0}u_0\delta(s) \quad \Rightarrow \quad (1 - J\beta)v_R^{[rel]}(t) = \int_{-\infty}^{\infty} ds\, \frac{\dot{u}(t-s)}{u_0}G(s)$$

It follows that

[5.12] $$\begin{cases} (1 - J\beta)v_{Rx}^{[rel]}(t) = \omega_h G_0 \int_0^{t_{ff}} ds\, (e^{-\gamma_R t_{ff}} - e^{-\gamma_R s})\sin\omega_h(t-s) = \\ = G_0 \int_0^{t_{ff}} ds\, (-\gamma_R e^{-\gamma_R s})\cos\omega_h(t-s) - G_0(e^{-\gamma_R s} - e^{-\gamma_R t_{ff}})\cos\omega_h(t-s)\Big|_0^{t_{ff}} = \\ = u_0\cos\omega_h t - \gamma_R G_0 \int_0^{t_{ff}} ds\, e^{-\gamma_R s}\cos\omega_h(t-s) \end{cases}$$

In the limit for vanishing $t_{ff}$ or $\gamma_R$, the response velocity does indeed coincide with the driver's velocity:

[5.13] $$(1 - J\beta)\lim_{t_{ff}\downarrow 0} v_{Rx}^{[rel]}(t) = u_0\cos\omega_h t = u(t)$$

Performing the integral of Eq. 5.10, the velocity response becomes

[5.14] $$(1 - J\beta)v_{Rx}^{[rel]}(t) = v_c\cos\omega_h t - v_s\sin\omega_h t = \sqrt{v_c^2 + v_s^2}\cos(\omega_h t + \psi_{hR})$$

with the definitions

$$[5.15] \quad \begin{Bmatrix} v_c - u_0 \equiv G_0 \gamma_R \dfrac{\gamma_0 e^{-\gamma_R t_{ff}} - \gamma_R}{\omega_h^2 + \gamma_R^2} \\ v_s \equiv G_0 \gamma_R \dfrac{\omega_h - \gamma_0 e^{-\gamma_R t_{ff}}}{\omega_h^2 + \gamma_R^2} \end{Bmatrix} \quad \& \quad \begin{Bmatrix} \tan \psi_{hR} \equiv \dfrac{v_s}{v_c} \\ \gamma_0 \equiv \gamma_R \cos \omega_h t_{ff} - \omega_h \sin \omega_h t_{ff} \end{Bmatrix}$$

The 3D plots below illustrate the sphere's velocity response as a function of $t_{ff}$ and $\gamma_R$.

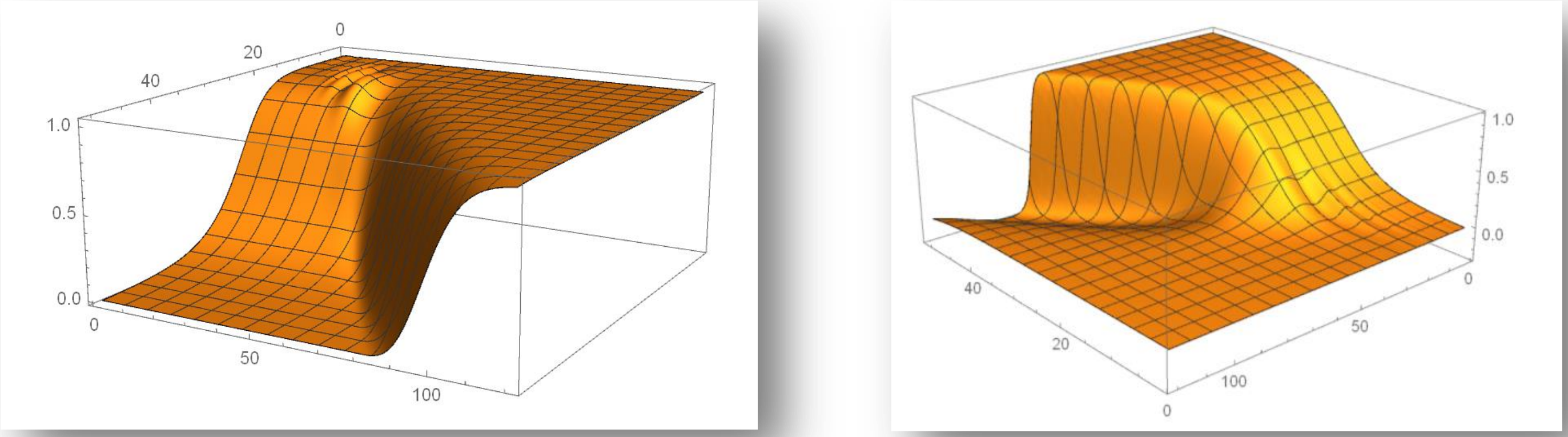

***Figure 5.1: Sphere motion in a bath of seeds at high temperature.*** *The vertical axis of the leftmost figure represents the amplitude of the scaled velocity* $(1 - J\beta)v_R^{[rel]}(t)$ *(see Eqs. 5.12 and 5.13). The horizontal axis reaching until 50 represents* $5\log[\tilde{\gamma}_R]$*, corresponding to a maximum value of* $\gamma_R = e^{10} s^{-1}$*, while that going to 120 represents* $-\log[\tilde{t}_{ff}]$*, corresponding to a minimum value of* $t_{ff} = e^{-12} s$*. Here the superposed wiggles indicate the corresponding dimensionless quantities with respect to 1 (reciprocal) second. The rightmost graph has identical horizontal axes. The vertical axis represents the phase* $\psi_{hR}$ *in units of* $\frac{1}{2}\pi$*. The phase never exceeds* $\frac{1}{2}\pi$*, and dives to negative values only in a small trough for* $\tilde{\gamma}_R \tilde{t}_{ff}^{0.182} = 2.6\%$*. The trough coincides with the fast dive in amplitude along the same logarithmic line. The perspective of the two graphs is different for better appreciation of the details. However, the* $t_{ff}$ *wiggles in the graphs coincide, and the horizontal axes fall on top of one another when the leftmost graph is turned 90° along the negative z-axis.*

Alas, the here presented theory is not able to determine the value of $\beta$, the ratio between the angular and linear velocities (cfr. Eq. 5.5). If that ratio is unity, there is no power loss due to friction of the sphere with respect to the substrate, as the sphere never slips – a rather improbable scenario. For ratios beyond unity, the sphere slips forward; it slips backward for ratios below unity.

# 1.6. Temperature of the Seed Bath

The simplicity of the previous Section's model is apparent from the fact that

(i) the amount and direction of sphere slipping is constant over the driving period (see Eq. 5.5);

(ii) a single frequency $\gamma_R$ describes the viscosity of the seed bath for both velocities, angular and linear.

These two simplifications are essential for visualizing the major trends, as in Fig. 5.1.

Before presenting an estimation of the temperature of the seed bath, we first discuss whether the seeds are likely to stick to the tray. In Lozano's experiment [24], at a seed concentration of 60% the average seed distance is 1.76 *mm*, implying a surface-to-surface separation of 0.70 *mm*. As a full swing of a single seed requires 1.2 *mm* (see Eq. 6.20) plus two seed radii (totaling at 2.26 *mm*), on average a seed can only perform 31% of its natural response. That is to say, at 60% concentration the seeds need to suffer many impacts before they can come to a natural stall. This leads us to the simplifying *Ansatz* that, above 50% concentration, seeds never stick to the tray. The relative velocity of a single seed then follows from [see Eqs. 3.6 and 3.7 in ref. 30], yielding

[6.1] $$v_{rx1}^{[rel]}(t) = v_{rx1}^{[abs]}\cos(\omega_h t - \phi_h) - u_0 \cos\omega_h t = -v_{hr}^{[rel]}\cos(\omega_h t - \phi_{hr}^{[rel]})$$

with the definitions

[6.2] $\kappa \equiv \frac{\omega_h u_0}{\mu_{dyn} g}$ & $\begin{Bmatrix} f_c \equiv 1 - \cos\kappa \\ f_s \equiv \kappa - \sin\kappa \end{Bmatrix}$ & $\begin{Bmatrix} \tan\phi_h \equiv \frac{f_s}{f_c} \\ f_h \equiv \sqrt{f_c^2 + f_s^2} \end{Bmatrix}$ & $v_r^{[abs]} \equiv \frac{2 f_h}{\kappa^2} u_0$

and

$$[6.3] \quad \begin{Bmatrix} \tan\phi_{hr}^{[rel]} \equiv \frac{v_r^{[abs]} \sin\phi_h}{u_0 - v_r^{[abs]} \cos\phi_h} \\ (v_{hr}^{[rel]})^2 \equiv (v_r^{[abs]} \sin\phi_h)^2 + (u_0 - v_r^{[abs]} \cos\phi_h)^2 \end{Bmatrix}$$

The total slider's excursion amplitude results upon integrating Eq. 6.1:

$$[6.4] \quad x_r^{[rel]}(t) = -\frac{v_{hr}^{[rel]}}{\omega_h} \sin(\omega_h t - \phi_{hr}^{[rel]})$$

Substitution of the experimental values in Eqs. 6.1 through 6.4 yields:

$$[6.5] \quad \begin{Bmatrix} \overline{[x_r^{[rel]}(t)]^2} = \frac{1}{2}(\frac{v_{hr}^{[rel]}}{\omega_h})^2 \xrightarrow{\text{[App.III]}} \frac{1}{2}(\frac{90\, mms^{-1}}{24\pi\, s^{-1}})^2 \approx 0.7\, mm^2 \\ \overline{x_h(t)^2} \xrightarrow{\text{[App.III]}} \frac{1}{2} x_{tray}^2 \approx 1.51 mm^2 = (1.2 mm)^2 \end{Bmatrix}$$

The seed's average squared displacement is about half that of the driver, and more importantly, still twice the maximum value reported by Lozano et al. for the spheres [24, Supporting Information, Fig. 2b]. This confirms the *Ansatz* that, at least above 60% concentration, seeds never stick to the tray.

Now we turn to the main purpose of this Section. The simplest way to estimate the temperature is to calculate the average longitudinal kinetic energy of a single seed over one oscillation period:

$$[6.6] \quad k_B T_{rx1} = \overline{E_{kin-rx1}(t)} = \tfrac{1}{2} m_r \overline{[v_{rx1}^{[rel]}(t)]^2} = m_r u_0^2 [(\frac{f_h}{\kappa^2})^2 - \frac{f_h}{\kappa^2} \cos\phi_h + \tfrac{1}{4}]$$

In case of ref. 24, $\kappa = \mu_{dyn}^{-1}$, and $\omega_h u_0 = g$, whence this simplifies to

$$[6.7] \quad k_B T_{rx1} = m_r u_0^2 \chi(\mu_{dyn}) \xrightarrow{[24]} 2 \times 10^{-9}\, J \cdot \chi(\mu_{dyn})$$

with

$$[6.8] \quad \chi(\mu_{dyn}) \equiv 1/4 + \mu_{dyn}^{2}\cos\mu_{dyn}^{-1} - 2\mu_{dyn}^{3}\sin\mu_{dyn}^{-1} + 2\mu_{dyn}^{4}(1-\cos\mu_{dyn}^{-1})$$

Due to inter-seed collisions, the temperature of the seed bath decreases with increasing concentration, dropping to 0 K at full tray filling (assuming the idealized absence of bilayer formation). The simplest dependence on concentration, respecting the boundary conditions of concentration, is

$$[6.9] \quad k_B T_{rx} = (1-C_r)^q k_B T_{rx1} = (1-C_r)^q \, 1/2\, m_r v_{rx1}^2$$

for $q$ = 1. Here we introduced the notational shorthand

$$[6.10] \quad v_{rx1}^2 \equiv \overline{[v_{rx1}^{[rel]}(t)]^2}$$

## 1.7. Transverse and Longitudinal Temperatures

The tray imparts longitudinal momentum onto the seeds, which bounce off the sphere in all directions. In this Section, we assume that the scattering efficiency of the moving sphere is equal to that of a sphere stuck to the tray. This will slightly underestimate the transverse temperature.

Seed concentration or filling fraction ( $C_r$ ), heart-to-heart nearest neighbor distance in a square crystalline organization ( $d$ ) and radius (***r***) of the seeds are related as

$$[7.1] \quad C_r \equiv N_r \frac{\pi r^2}{(LW)_0} = (\frac{2r}{d})^2$$

As the seeds have a kidney form, a radius is not enough to characterize their shape, let alone their closest packing configuration. Consequently, the seed radius is a quantity *defined* by Eq. 7.1 as a function of concentration, rather than *vice versa*.

Suppose, a quarter circle scatters incoming point dot-like seeds in two dimensions, as depicted in Fig. 7.1.

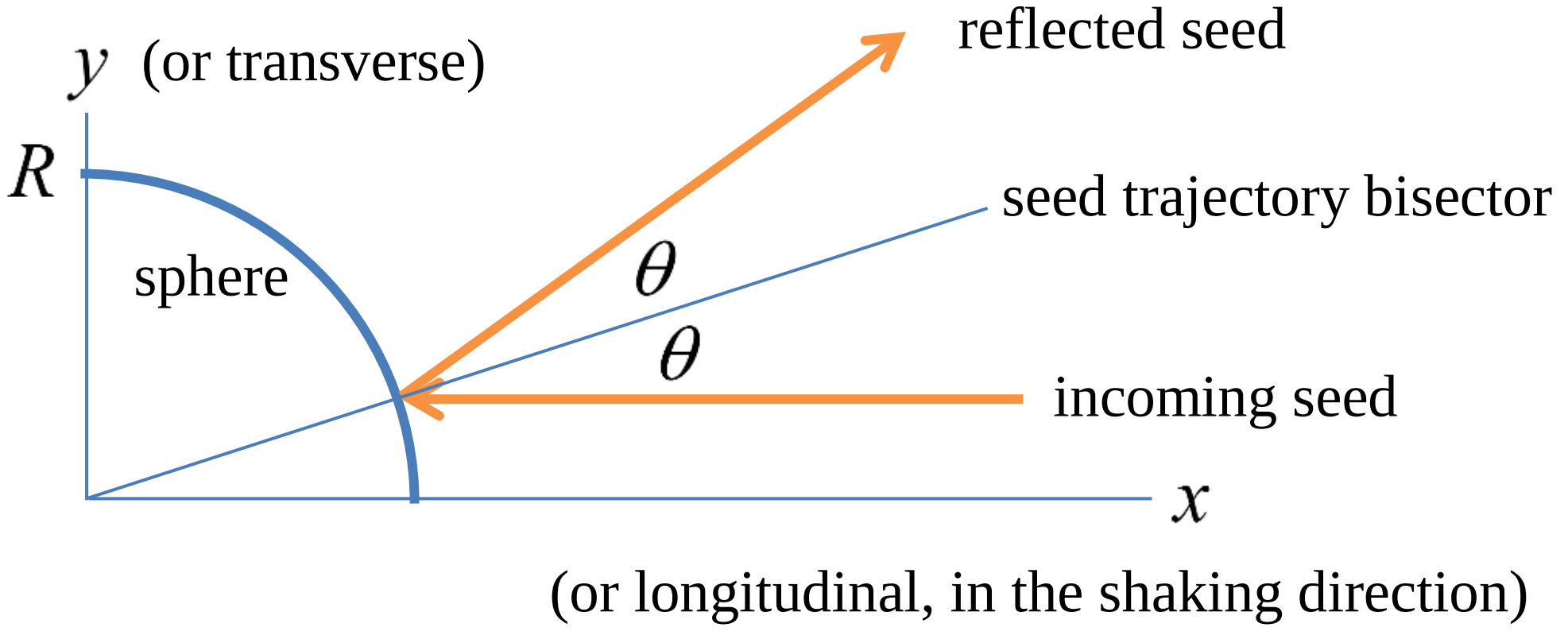

***Figure 7.1: 2D seed scattering off a sphere.*** *Both x and y–axes are in the horizontal plane. For simplicity of the calculations, the seed behaves like a point particle.*

An elastically reflected seed has a transverse velocity

[7.2] $$v_{ry1} = v_{rx1} \sin 2\theta$$

for $0 \le \theta \le \frac{1}{2}\pi$ . The average square transverse velocity results from

[7.3] $$v_{ry1}^2 = \frac{2}{\pi} v_{rx1}^2 \int_0^{\pi/2} d\theta \, \sin^2 2\theta = \frac{1}{2} v_{rx1}^2$$

The impact rate of seeds onto a sphere yields the temperature ratio, for the conditions of ref. 24,

[7.4] $$\frac{T_{ry}}{T_{rx}} = \frac{1}{2}(1 - \frac{\gamma_W}{\omega_h}) \frac{(2R)^2}{WL} < 7 \times 10^{-5}$$

Here, $\gamma_W$ designs the rate at which transverse kinetic energy converts into longitudinal kinetic energy, due to reflection off the transverse tray borders followed by transverse inter-seed collisions. We will omit this factor in further estimations, as $\gamma_W \ll \omega_h$. This fixes the longitudinal (*x*) and transverse (*y*) seed bath (*r*) temperatures as

[7.6] $$\begin{Bmatrix} T_{rx} \approx (1-C_r)T_{rx1} \\ T_{ry-1sphere} < 7\times 10^{-5}(1-C_r)T_{rx1} \end{Bmatrix}$$

Clearly, the transverse temperature plays no significant role, and can be set to zero for all practical applications. Obviously, this drastically changes for high sphere concentrations.

## 1.8. Brownian Motion of the Sphere

Using Eqs. 6.7 through 6.9, the longitudinal diffusion of the sphere is

[8.1] $$D_{Rx} = \frac{k_B T_{rx}}{m_R \gamma_v(C_r)} = \chi(\mu_{dyn})(1-C_r)\frac{k_B T_{rx1}}{m_R \gamma_v(C_r)}$$

The reciprocal frequency $\gamma_v^{-1}$ represents the time it takes for a moving sphere, in a seed bath of 0 K, to reduce its initial velocity by a factor *e*. From Fig. 7.1 one can see that, for a longitudinally impinging seed with velocity *v* at transverse height $y = R\cos\theta$ elastic momentum balance requires, along the bisector $\hat{\theta}$ of the incoming and outgoing seed path,

[8.2] $$(m_R \Delta v_{Rx}\hat{\theta} + 2m_r v_{rx}\cos\theta)\hat{\theta} = 0$$

Hence, along the longitudinal axis, the seed's velocity decrease due to a single sphere-seed collision is

[8.3] $$(m_R \Delta v_{Rx} + 2m_r v_{rx}\cos^2\theta)\hat{x} = 0$$

The 2D-average over all impinging heights is

[8.4] $$\overline{\cos^2\theta} = \frac{1}{R}\int_0^R dy\,\cos^2\theta = \frac{1}{R}\int_0^R dy\,(1-\sin^2\theta) = \frac{1}{R}\int_0^R dy\,[1-(\frac{y}{R})^2] = \tfrac{2}{3}$$

In Eq. 8.3, $v_{rx}$ represents the approach velocity of sphere and seed along the longitudinal direction. Since the sphere moves, and the seeds stand still (bath of 0 K: see Eq. 4.9), one obtains

[8.5] $$v_{Rx}^{[aft]} = \frac{1}{3}(1+\frac{4m_r}{m_R+m_r})v_{Rx}^{[bef]}$$

The time needed for 100 collisions at a given seed concentration is

[8.6] $$\gamma_v^{-1} = \frac{100d^2}{Rv_{in}} = C_r^{-1}\frac{100(2r)^2}{Rv_{in}} \xrightarrow{[24]} 2s\cdot C_r^{-1}$$

The sphere's longitudinal diffusion becomes

[8.7] $$\left\{\begin{array}{l} D_{Rx}(C_r) = 2sC_r^{-1}\frac{k_B T_{rx}}{m_R} \xrightarrow{[24]} 250\,mm^2 s^{-1}\cdot\chi(\mu_{dyn})\frac{(1-C_r)^q}{C_r} \\ D_{Rx}(0.6) = 416\cdot(0.4)^q\,mm^2 s^{-1}\cdot\chi(\mu_{dyn}) = 27\,mm^2 s^{-1} = 0.36\,\omega_h\,mm^2 \end{array}\right\}$$

This condition is satisfied for

[8.8] $$\zeta(q)\cdot\chi(\mu_{dyn}) = 1 \quad \& \quad \zeta(q) \equiv 15.41\,(0.4)^q$$

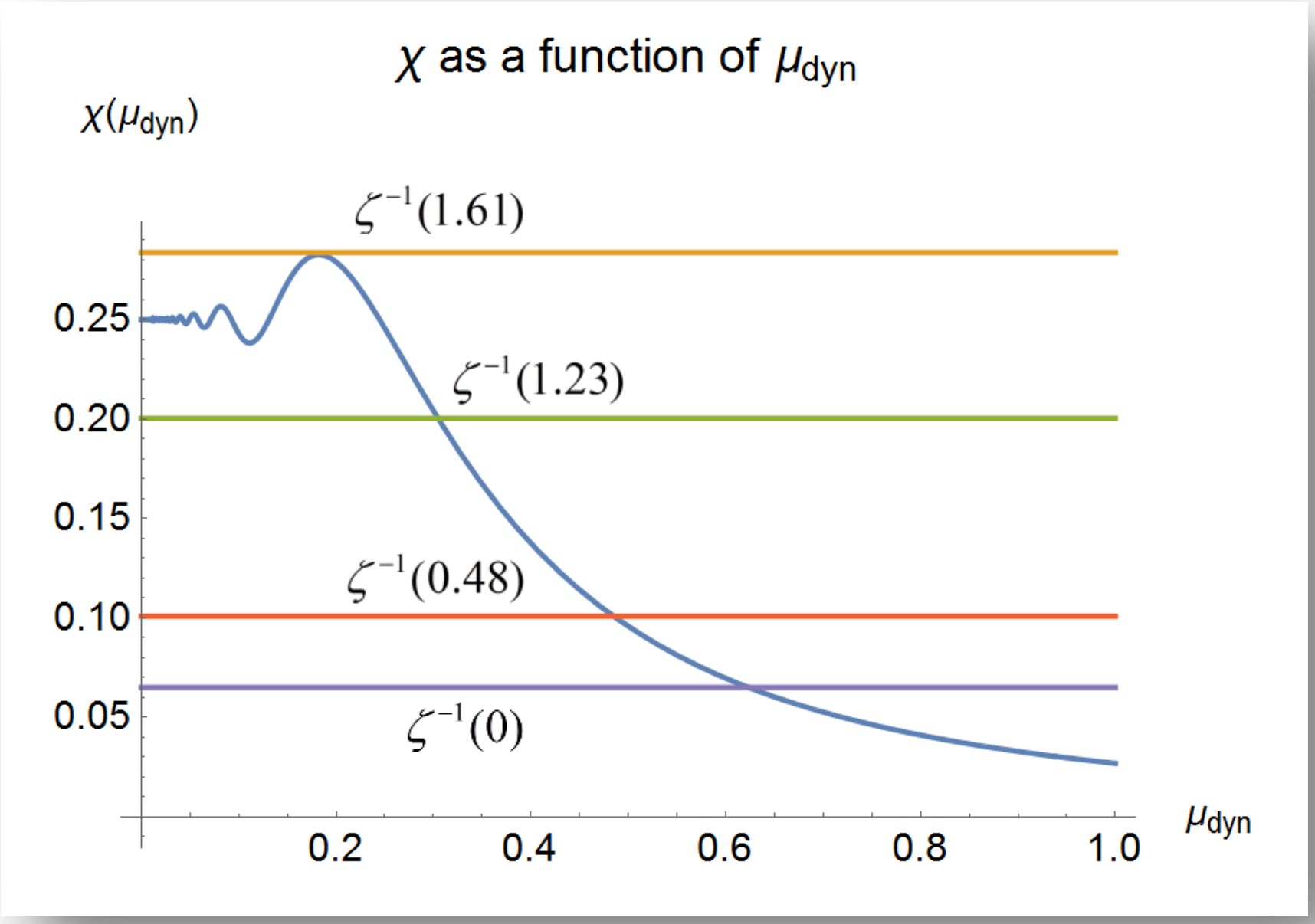

***Figure 8.1: Friction Dependence of the dimensionless function*** $\chi(\mu_{dyn})$**.** *$\chi(\mu_{dyn})$ peaks at 28.5%, where the dynamic friction coefficient equals 20%. Four horizontal lines mark the values of $\zeta^{-1}(q)$ for selected values of q. Taking Lozano's [24] measurement point at 60% seed concentration, as a granted match, our model is not able to account for seed dynamic friction coefficients beyond 62%.*

One might think that the model is able to fit an elephant. However, the dynamic friction coefficient of a seed on the tray is a measurable quantity: it is not a fit parameter. Once the coefficient is measured, there remains but a single fit parameter ($q$) to fit theory to experiment. Given Fig. 8.1, this should only be possible for $0 < q < 1.61$ (see Fig. 8.1). For the four $q$-values displayed in Fig. 8.1, we will now present the theoretical predictions of the concentration dependence of the diffusion coefficient, along with the experimentally measured points in reference 24.

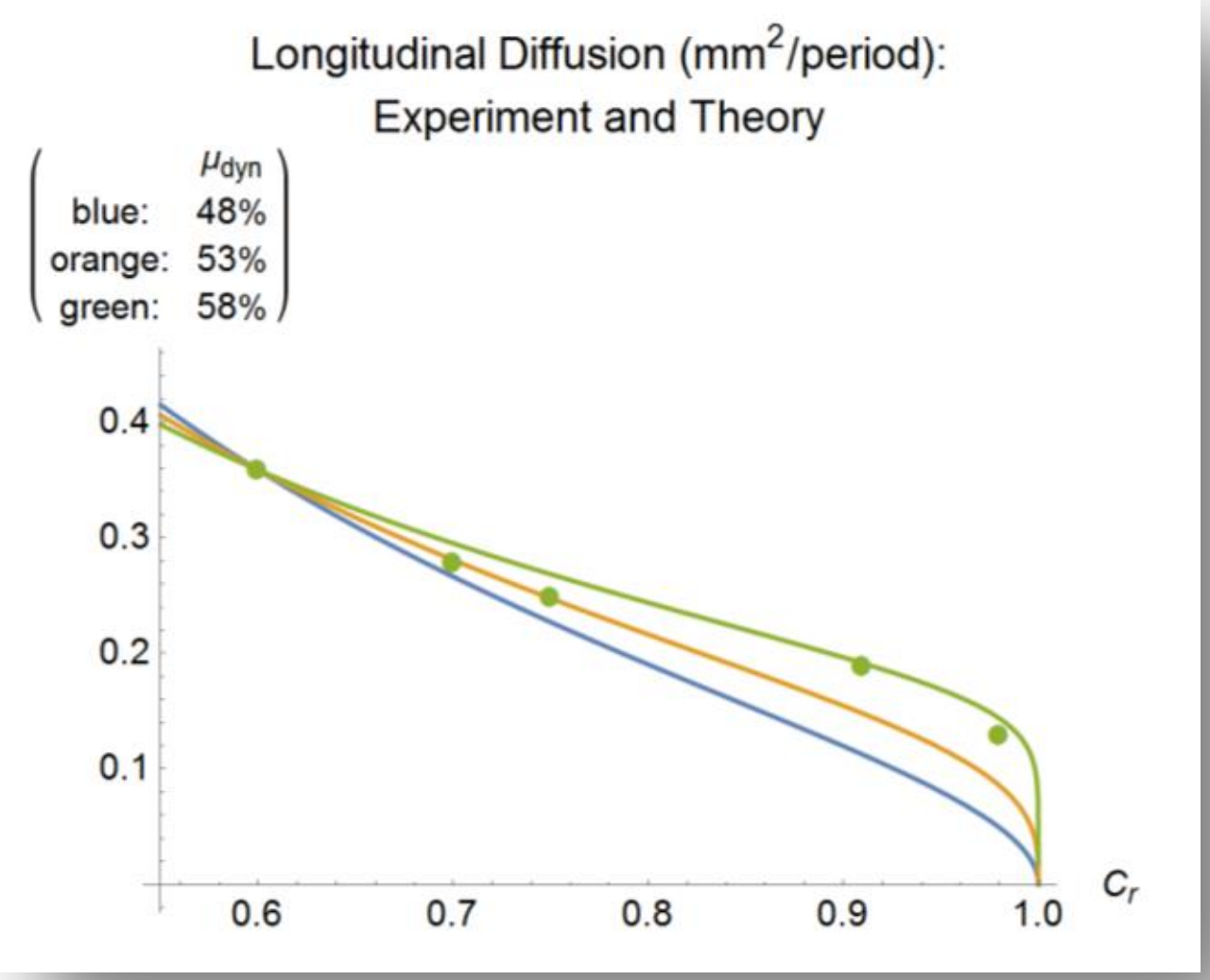

***Figure 8.2: Concentration Dependence of the longitudinal diffusion,*** *experiment [24] and theory (for three different friction values). All three curves pass through the measurement at lowest concentration.*

As mentioned above, we attach more importance to the experimental data at low concentration. On one hand, seed bilayer formation and sphere jumping (on top of a seed monolayer) might play a role at concentrations beyond 85%. On the other hand, if the mentioned effects do not occur, the high temperature limit (one of the *Ansätze* of our model) is not satisfied at the highest concentrations. From Fig. 8.2, one may deduce that 53% < $\mu_{dyn-seed}$ < 58%.

The measurements show that the transverse diffusion has a concentration-independent value of

[8.9] $$D_{rx} = 0.05\omega_h \, mm^2 = 3.8\, mm^2 s^{-1}$$

A departure from that value occurs only at the highest concentrations [24], but there the high longitudinal temperature condition is not satisfied. This confirms our intuition that all transverse motion of the sphere is due, not to the seeds' transverse impacts (see Eq. 7.6), but to their longitudinal impacts. At lower seed concentrations, one may graphically represent the seeds' collective motion as a viscous hydrodynamic flow superposed onto the the more prominent harmonic motion. If vortices establish both transversally above and below the sphere, the problem of explaining the transverse experimental data [24] becomes a complex hydrodynamic one, with a strong influence of the vertical tray borders on the seed friction. That goes quite beyond the scope of this Chapter.

# Chapter II:

# Response of a Sphere in a Harmonically Shaken Bath

## 2.1. Single Sphere: Seed Baths' Configuration Entropy

Every poppy seed in physical contact with a two-dimensional (horizontal) tray has five degrees of freedom: two positional degrees, two velocity degrees, and one rotational degree (around the vertical *z*-axis, with *x* the horizontal driving or longitudinal axis, and *y* the horizontal transverse axis). The heavy sphere has two more: it is able to rotate around the two horizontal axes, too.

Consider a monolayer seed bath with a surface concentration 50%< $C_r$ < 100%, and a single heavy sphere immersed into it. Assume the sphere's surface to be negligible with that of the tray. In this paper we assume the experimental parameters chosen by Lozano et al. [24]. The Appendix of Chapter III of this article resumes all their system parameters. For the specific condition $g = u_0\omega_h$, with $g$ the earth's gravity acceleration, $u_0$ the velocity amplitude of the harmonic driver, and $\omega_h = 2\pi\nu_h$ the tray's driving frequency, and a harmonic oscillation frequency $\nu_h = 12 Hz$, the latter is always smaller than the impact frequency of seeds onto a the sphere: $30\,Hz < \gamma_{impact} = 50\ C_r\ Hz < 50\,Hz$ .

Moreover, as shown in Eq. 6.5 of Ref. 30, beyond 60% concentration ($1mm < d \equiv 2r\,C_r^{-1/2} < 1.4\,mm$), the driver amplitude ( $x_{h-tray} = 12.7\,mm$ ) by far exceeds the inter-seed distances. These two magnitude comparisons allow one to consider the longitudinal energy equipartition theorem to apply at all times. The adjective "longitudinal" refers to the *x*-direction or oscillation direction. The *y*- or "transverse" direction is in the horizontal plane, too, though perpendicular to the oscillation direction.

The sphere's 2D-contribution to the position entropy,

[1.1] $$S_{R-stat}^{[2D]}(N_r) = k_B \ln \frac{LW - N_r \pi r^2}{\pi R^2},$$

is in general negligible with that of the seeds:

[1.2] $$S_{r-stat}^{[2D]}(N_r) = N_r k_B \ln \frac{LW - \pi R^2}{\pi r^2}$$

with $N_r$ the number of seeds, and $L$ and $W$ the length and width of the tray confining the seed bath, respectively. The static thermodynamic expressions 1.1 and 1.2 represent the positional equilibrium entropy. Once the tray starts moving harmonically, regions of compaction and dilution (at the longitudinal front and back of the sphere, respectively) appear, which reduce the seeds' position entropy to the dynamical value $S_{dyn}^{[1D]}(t, N_r)$. In order to determine the 2D-logarithmic argument, we first develop an expression valid in a pseudo-1D tray, which has a width $W = 2R$ able to contain the heavy sphere. The static 1D seed entropy is, for $N_r$ indistinguishable seeds,

[1.3] $$\left\{ \begin{array}{l} \dfrac{S_{r-stat}^{[1D]}(N_r)}{k_B} = \ln \begin{pmatrix} N_{tot} \\ N_r \end{pmatrix} = \ln \dfrac{N_{tot}!}{(N_{tot} - N_r)!\, N_r!} = \ln N_{tot}! - \ln(N_{tot} - N_r)! - \ln N_r! = \\ \approx N_{tot} \ln N_{tot} - (N_{tot} - N_r)\ln(N_{tot} - N_r) - N_r \ln N_r \end{array} \right\}$$

The total number of accessible spots given by

[1.4] $$N_{tot}^{[1D]} \equiv \frac{L - 2R}{2r}$$

Assume the sphere to have an equal number of seeds to its right as to its left, and the two characteristic lengths to be equal: $L_{high} = L_{low} \equiv L_{char}$. The available length for the seeds is always larger than the sum of the two characteristic lengths: $L - 2R \geq 2L_{char}$. The pseudo-1D-dynamic entropy $S_{r-dyn}^{[1D]}(u_0, \omega_h t, N_r)$ must satisfy the following conditions:

(i) $$\lim_{u_0 \downarrow 0} S_{r-dyn}^{[1D]}(u_0, \omega_h t, N_r) = S_{r-stat}^{[1D]}(N_r)$$

(ii) $$\lim_{u_0 \uparrow \infty} S_{r-dyn}^{[1D]}(u_0, \omega_h t, N_r) = k_B \approx 0$$

The simplest 1D-entropy fulfilling the above conditions is

[1.5] $$\Delta S_r^{[1D]}(u_0,\omega_h t,N_r) \equiv S_{r-stat}^{[1D]}(N_r) - S_{r-dyn}^{[1D]}(u_0,\omega_h t,N_r) = \frac{S_{r-stat}^{[1D]}(N_r)}{1+\dfrac{\Delta v_{Rr}(\omega_h t)}{u_0}}$$

with the differential (sphere-in-bath minus bath alone) velocity

[1.6] $$\Delta v_{Rr}(\omega_h t) \equiv v_{Rx}(\omega_h t) - v_{rx}(\omega_h t)$$

We omitted the superscripts [rel] or [abs] on the two velocities on the RHS of Eq. 1.6, because the difference velocity is independent of the Galilean reference frame of the two RHS velocities.

The same Eq. 1.5 holds for two dimensions, too.

[1.7] $$\left\{ \begin{array}{l} \dfrac{\Delta S_r^{[2D]}(u_0,\omega_h t,N_r)}{S_{r-stat}^{[2D]}(N_r)} \equiv 1 - \dfrac{S_{r-dyn}^{[2D]}(u_0,\omega_h t,N_r)}{S_{r-stat}^{[2D]}(N_r)} = \dfrac{1}{1+\dfrac{\Delta v_{Rr}(u_0,\omega_h t,N_r)}{u_0}} \\ \dfrac{\Delta v_{Rr}(\omega_h t)}{u_0} << 1: \quad \dfrac{\Delta S_r^{[2D]}(u_0,\omega_h t,N_r)}{S_{r-stat}^{[2D]}(N_r)} \approx 1 - \dfrac{\Delta v_{Rr}(u_0,\omega_h t,N_r)}{u_0} \end{array} \right\}$$

Section 3 of this Chapter presents an example using realistic experimental parameters.

## 2.2. The Sphere's Harmonic Response in a Seed Bath

The sphere's unhindered (i.e., by the seeds) velocity deceleration on a stalled substrate fulfills,

[2.1] $$\dot{v}_R^{[dec]} + \zeta_\mu^{-1} g \,\mathrm{sgn}[v_R^{[dec]}] \propto \delta(t)$$

where $\zeta_\mu^{-1} \equiv \frac{\mu_{dyn}}{1 - J\beta}$ represents the roller's "generalized sliding friction". Due to the continuous seed-bombardment of the seed-immersed sphere, Eq. 2.1 needs a deceleration term, describing the sphere's momentum loss due to seed collisions. It depends crucially on sphere-seed mass ratio, temperature, and seed concentration.

We start investigating the deceleration of a sphere when unleashed frictionless (with the substrate) into a 0 K seed bath, using the conventions of Fig. 7.1 of Chapter I. For a given transverse displacement, $\delta y \equiv R\cos\theta \neq 0$ (note that we therefore do *not* describe a head-on collision), we now choose the initial velocities of sphere and seed as

[2.2] $$\left\{ \begin{matrix} v_{Rx}^{[bef]} > 0 \\ v_{Ry}^{[bef]} = v_{rx}^{[bef]} = v_{ry}^{[bef]} = 0 \end{matrix} \right\}$$

Due to momentum conservation, one obtains the final velocities

[2.3] $$\left\{ \begin{matrix} v_{Rx}^{[aft]} = \dfrac{2m_r \mathbf{v}_{r\theta}^{[bef]} + (m_R - m_r)\mathbf{v}_{R\theta}^{[bef]}}{m_R + m_r} \cdot \hat{\mathbf{x}} = \dfrac{m_R - m_r}{m_R + m_r} v_{Rx}^{[bef]} \cos^2\theta \\ v_{rx}^{[aft]} = \dfrac{2m_R \mathbf{v}_{R\theta}^{[bef]} - (m_R - m_r)\mathbf{v}_{r\theta}^{[bef]}}{m_R + m_r} \cdot \hat{\mathbf{x}} = 2\dfrac{m_R}{m_R + m_r} v_{Rx}^{[bef]} \cos^2\theta \end{matrix} \right\}$$

The light seed shoots away in forward direction (that of the moving sphere). Per collision, the heavy sphere loses about a third of its initial velocity, (see Eq. 8.5 of Chapter I). Assume the density of seeds in front of the sphere is constant. Then the collision rate becomes proportional to the sphere velocity and concentration:

[2.4] $$\gamma_{coll} = \frac{2R}{d}\frac{v_{Rx}}{d} = \frac{C_r R v_{Rx}}{2r^2}$$

it follows that the sphere's velocity decays hyperbolically

$$[2.5] \quad \lambda^{-1} \equiv \frac{C_r R}{3r^2}\left(1 - \frac{2m_r}{m_R + m_r}\right) = \frac{C_r R}{3r^2}\frac{m_R - m_r}{m_R + m_r} \quad \Rightarrow \quad v_{Rx}^{-2}\frac{dv_{Rx}}{dt} = -\frac{dv_{Rx}^{-1}}{dt} = \lambda^{-1}$$

with a curvature determined by the characteristic length

$$[2.6] \quad \lambda \equiv \frac{3r^2}{C_r R}\frac{m_R + m_r}{m_R - m_r} \quad \Rightarrow \quad v_{Rx}(t) = \frac{\lambda}{t_0 - t}$$

Translated into typical numbers [24], one finds

$$[2.7] \quad \lambda C_r \equiv \frac{3r^2}{R}\frac{m_R + m_r}{m_R - m_r} \xrightarrow{[App.III]} 1.15\,mm$$

Consequently, using Eq. 2.6 and the initial condition, $v_{Rx}(0, C_r) = u_0$, one obtains the average collision time

$$[2.8] \quad \begin{Bmatrix} t_0 \equiv \dfrac{\lambda}{u_0} \xrightarrow{[App.III]} C_r^{-1} 8.8\,ms \\ \omega_h t_0 = \dfrac{8.8\,ms}{C_r} = \dfrac{0.6653}{C_r} \approx \dfrac{2}{3C_r} \end{Bmatrix}$$

Combined with Eq. 2.1, one obtains the equation of motion for a friction-decelerated sphere in a 0 K seed bath

$$[2.9] \quad \dot{v}_R^{[dec]}(t, C_r, \zeta_\mu) + \frac{u_0}{t_0 - t} + \zeta_\mu^{-1} g \,\mathrm{sgn}[v_R^{[dec]}(t, C_r, \zeta_\mu)] = u_0 \delta(t)$$

with the solution

$$[2.10] \quad \frac{v_R^{[dec]}(t, C_r, \zeta_\mu)}{u_0} = [1 + \ln\frac{t_0 - t}{t_0} - \zeta_\mu^{-1}\omega_h t]\theta(t)\theta(t_{ff} - t)$$

This implies that the initial deceleration is

$$[2.11] \quad -\frac{\dot{v}_R^{[dec]}(0^+, C_r, \zeta_\mu)}{u_0} = (\tfrac{3}{2} C_r + \zeta_\mu^{-1})\omega_h \equiv \frac{1}{t_{ff\,0}} \quad \Rightarrow \quad \omega_h t_{ff\,0} = (\tfrac{3}{2} C_r + \zeta_\mu^{-1})^{-1}$$

Hence, the steepness of the deceleration velocity decreases linearly both with seed concentration $C_r$ and with the friction $\zeta_\mu^{-1}$ (see Eq. 2.1). We shall call its reciprocal value, $\zeta_\mu$, a measure of glibness. The steepness of the initial deceleration velocity agrees with three common-sense requirements:

(i) the higher the concentration, the faster the sphere's initial velocity decreases (assuming a spatially homogenous distribution of seeds);

(ii) the higher the seed friction with the substrate, the lower the seeds' average velocity, and the higher their inertia;

(iii) the higher the product of angular momentum $J$ and of the sphere's friction parameter $\beta$, the faster the sphere's initial velocity decreases.

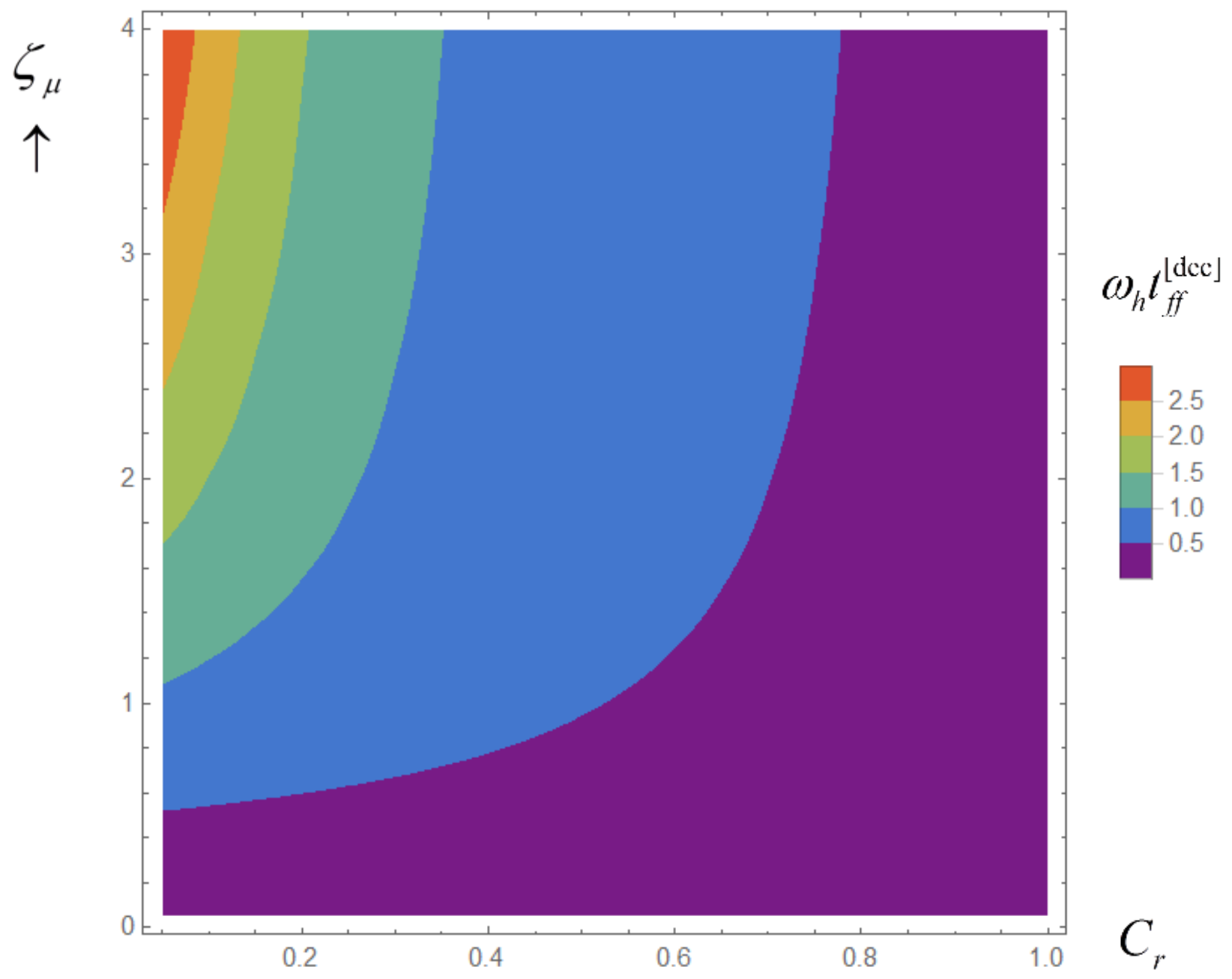

***Figure 2.1: Contour Plot of dimensionless free flight time** $\omega_h t_{ff}$ (see Eqs. 2.15 or 2.16) as a function of concentration $C_r$ on the x-axis, and of glibness (inverse friction) $\zeta_\mu$ on the y-axis. The lowest free flight time occurs at (1,0) and the highest at (0,4) for the coordinate pair $(C_r, \zeta_\mu)$.*

Fig 2.1 confirms the obvious notions that the free flight time has

(i) a minimum value for maximum concentration (due to the high rate of collisions of the sphere with the seeds) in combination with the lowest glibness (highest friction);

(ii) a maximum value for zero concentration in combination with maximum glibness; in this case there are no slowing-down collisions with the seeds, and the friction of the sphere at velocity reversal is minimal.

Most importantly, we now have all ingredients to calculate the sphere's response to harmonic driving. When the tray oscillates at a velocity *u(t)*, the harmonic response becomes the following convolution

[2.12] $$\left\{\begin{aligned} & v_{Rxh}^{[abs]}(t,C_r,\zeta_\mu) \equiv N^{-1}\int_0^{t_{ff}} ds\, u(t-s) v_R^{[dec]}(s,C_r,\zeta_\mu) \\ & N \equiv \left|\int_0^{t_{ff}} ds\, v_R^{[dec]}(s,C_r,\zeta_\mu)\right| \end{aligned}\right\}$$

or, in dimensionless variables, and omitting the superfluous absolute value:

[2.13] $$\left\{\begin{aligned} & \frac{v_{Rxh}^{[abs]}(\omega_h t,C_r,\zeta_\mu)}{u_0} \equiv M^{-1}\int_0^{\omega_h t_{ff}} dz\, \cos(\omega_h t - z) K(z,C_r,\zeta_\mu) \\ & M \equiv \int_0^{\omega_h t_{ff}} dz\, K(z,C_r,\zeta_\mu) \end{aligned}\right\}$$

with the kernel (dimensionless velocity decay) defined as

[2.14] $$K(z,C_r,\zeta_\mu) \equiv 1 + \ln(1 - \tfrac{3}{2} C_r\, z) - \zeta_\mu^{-1} z$$

The (first) zero of the velocity decay defines the free flight time

[2.15] $$K(\omega_h t_{ff},C_r,\zeta_\mu) = 0$$

For $0 < t < t_{ff}$ the velocity-decay is a strictly positive, monotonically decaying function of time. The dimensionless free flight times follow by expanding the logarithm around unity while solving for Eqs. 2.14 and 2.15. A third order approximation is enough to reproduce exactly Fig. 2.1:

[2.16] $$\omega_h t_{ff}^{[3]}(C_r,\zeta_\mu) \equiv \frac{2}{3C_r} - \zeta_\mu \mathrm{PL}\left[\frac{2}{3C_r\zeta_\mu} e^{\frac{2}{3C_r\zeta_\mu}-1}\right]$$

The definition of the product logarithm is $z = \mathrm{PL}[ze^z]$, just like the ordinary logarithm has $z = \ln[e^z]$. Using the expansion and choice of first coefficient, the normalization reads

$$\text{[2.21]} \quad \left\{ \begin{array}{l} M \equiv \int_0^{\omega_h t_{ff}} dz\, K(z, C_r, \zeta_\mu) = -\dfrac{2}{3C_r} Q \log Q - \dfrac{(\omega_h t_{ff})^2}{2\zeta_\mu} \\ 0 < Q(C_r, \zeta_\mu) \equiv 1 - \dfrac{3}{2} C_r \omega_h t_{ff}(C_r, \zeta_\mu) < 1 \end{array} \right\}$$

The lab-frame-referenced and dimensionless velocity response acts like the kernel in the harmonic-drive convolution:

$$\text{[2.22]} \quad \frac{v_{Rxh}^{[abs]}(t, C_r, \zeta_\mu)}{u_0} \equiv M^{-1} \int_0^{\omega_h t_{ff}} dz\, \cos(\omega_h t - z) K(z, C_r, \zeta_\mu) = V_h \cos(\omega_h t - \phi_h)$$

The expressions for amplitude and phase are so ugly that we gathered them in the appendix.

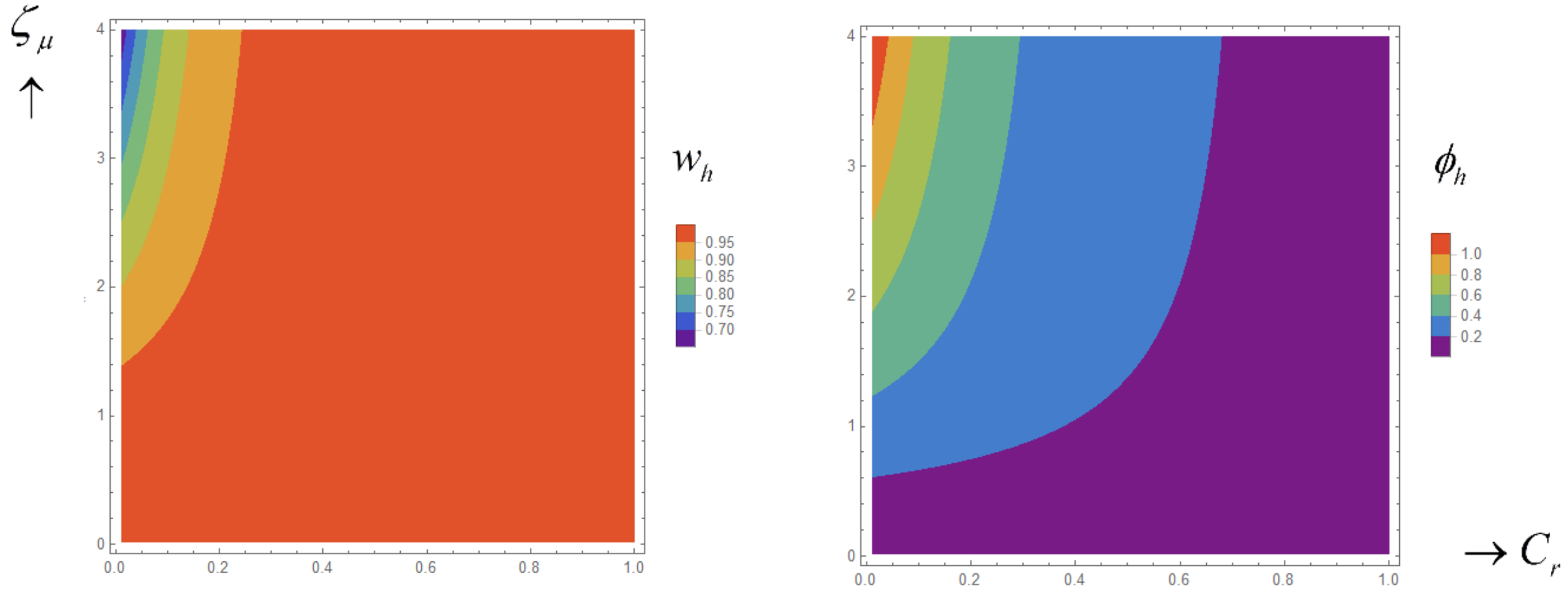

***Figure 2.2: 2D-contour plots of the sphere's amplitude and phase as a function of concentration and glibness.*** *In most of the parameter space, the laboratory-referenced amplitude is close to unity (i.e., the sphere is like stuck to the moving substrate), except at lowest concentration and friction. The phase vanishes with glibness, or with unit concentration.*

Important proviso: since we have described the sphere's motion assuming energy equipartition for the kinetic and rotational terms, the sphere's behavior at low concentration is questionable at least.

## 2.3. Average Linear Velocity of the Seed Bath Alone

The seed bath, when not too dilute, grants that a single seed practically never reaches spontaneous stalling, due to the continuous bombardment of colleague seeds. Hence, we choose as the seed bath's motion that of a slider in the non-stalling limit (see Chapter I, Eqs. 6.1 and 6.2):

[3.1] $$v_{rx}^{[abs]}(t) = v_{rx}^{[abs]} \cos(\omega_h t - \phi_h)$$

with the definitions

[3.2] $$\kappa \equiv \frac{\omega_h u_0}{\mu_{dyn} g} \quad \& \quad \begin{Bmatrix} f_c \equiv 1 - \cos\kappa \\ f_s \equiv \kappa - \sin\kappa \end{Bmatrix} \quad \& \quad \begin{Bmatrix} \tan\phi_h \equiv \dfrac{f_s}{f_c} \\ f_h \equiv \sqrt{f_c^2 + f_s^2} \end{Bmatrix} \quad \& \quad v_{rx}^{[abs]} \equiv \frac{2 f_h}{\kappa^2} u_0$$

This velocity only holds for poppy seeds outside the cones of influence of the spheres, defined in Chapter III. In order to specify the above expressions, we use experimental parameters:

[3.3] $$u_0 \xrightarrow{[App.III]} \frac{g}{\omega_h} \quad \Rightarrow \quad \kappa \to \mu_{dyn}^{-1} \quad \Rightarrow \quad \begin{Bmatrix} f_c \to 1 - \cos\mu_{dyn}^{-1} \\ f_s \to \mu_{dyn}^{-1} - \sin\mu_{dyn}^{-1} \end{Bmatrix} \quad \& \quad v_{rx}^{[abs]} \equiv 2\mu_{dyn}^2 f_h u_0$$

For $\mu_{dyn} \to 50\%$, $4(\frac{v_{rx}^{[abs]}}{u_0})^2 = f_h^2 = (1-\cos 2)^2 + (2-\sin 2)^2 = 3.2$ the seed's absolute velocity becomes

[3.4] $$\frac{v_{rx}^{[abs]}(t)}{u_0} \to 90\% \cos(\omega_h t - 17\%\pi)$$

The high amplitude (90%) implies that the seeds hardly (10%) move with respect to the tray, and they do so with a small phase delay. Define the velocity difference, the dimensionless help-velocities and angles

[3.5] $$\Delta v_{Rr}(\omega_h t) \equiv v_{Rx}^{[abs]}(\omega_h t) - v_{rx}^{[abs]}(\omega_h t) \equiv u_0 \Delta V_{Rr}$$

[3.6] $$\begin{Bmatrix} \Delta V_c \equiv (V_h - 90\%) \cos \tfrac{1}{2}\Delta\phi_h \\ \Delta V_s \equiv (V_h + 90\%) \sin \tfrac{1}{2}\Delta\phi_h \end{Bmatrix}$$

[3.7] $\begin{Bmatrix} \bar{\phi}_h \equiv \tfrac{1}{2}(\phi_h + 17\%\pi) \\ \Delta\phi_h \equiv \phi_h - 17\%\pi \end{Bmatrix}$ & $\begin{Bmatrix} \phi_h \equiv \bar{\phi}_h + \tfrac{1}{2}\Delta\phi_h \\ 17\%\pi \equiv \bar{\phi}_h - \tfrac{1}{2}\Delta\phi_h \end{Bmatrix}$

[3.8] $$\begin{Bmatrix} \Delta V_{Rr}(C_r, \zeta_\mu) \equiv \sqrt{\Delta V_c^2 + \Delta V_s^2} \\ \delta\phi(C_r, \zeta_\mu) \equiv \arccos(\frac{\Delta V_c}{\Delta V})\,\mathrm{sgn}(\arcsin\frac{\Delta V_s}{\Delta V}) \end{Bmatrix},$$

respectively, and one obtains just another harmonic response

[3.9] $$\begin{Bmatrix} \Delta V_{Rr}(\omega_h t) = V_h \cos(\omega_h t - \bar{\phi}_h - \tfrac{1}{2}\Delta\phi_h) - 90\% \cos(\omega_h t - \bar{\phi}_h + \tfrac{1}{2}\Delta\phi_h) = \\ = \begin{Bmatrix} V_h[\cos(\omega_h t - \bar{\phi}_h)\cos\tfrac{1}{2}\Delta\phi_h + \sin(\omega_h t - \bar{\phi}_h)\sin\tfrac{1}{2}\Delta\phi_h] + \\ -90\%[\cos(\omega_h t - \bar{\phi}_h)\cos\tfrac{1}{2}\Delta\phi_h - \sin(\omega_h t - \bar{\phi}_h)\sin\tfrac{1}{2}\Delta\phi_h] \end{Bmatrix} = \\ = \Delta V_c \cos(\omega_h t - \bar{\phi}_h) + \Delta V_s \sin(\omega_h t - \bar{\phi}_h) = \Delta V_{Rr} \cos(\omega_h t - \bar{\phi}_h - \delta\phi) \end{Bmatrix}$$

E.g., for the parameter couple $(C_r, \zeta_\mu) = (0.7, 1.6)$, one has

[3.10] $$\omega_h t_{ff}^{[3]} = 15\%\pi \quad \Rightarrow \quad \begin{Bmatrix} V_h = 0.99u_0 \\ \phi_h = 5\%\pi \\ \Delta\phi_h = -12\%\pi \end{Bmatrix} \quad \Rightarrow \quad \begin{Bmatrix} \Delta V_c \equiv 9\% \\ \Delta V_s \equiv -38\% \end{Bmatrix} \quad \Rightarrow \quad \begin{Bmatrix} \Delta V_{Rr} = 39\% \\ \delta\phi = -43\%\pi \\ \bar{\phi}_h = 11\%\pi \end{Bmatrix}$$

Hence, under typical circumstances, one finds a quite large discrepancy between the sphere's velocity and the seeds' velocity:

[3.11] $$\Delta V_{Rr}(u_0, \omega_h t, C_r \to 0.7, \xi_\mu \to 1.6) = 39\% \cos(\omega_h t + 31\%\pi)$$

The spheres lag behind the seeds for approximately 60°, with a difference amplitude of about $\tfrac{2}{5}$ the driver's. These are high numbers, which promise equally high entropic attraction effects.

# Chapter III: Attraction and Viscosity

## 3.1. Two Spheres: Entropic Force of Attraction

In this Section, we first consider the pseudo-2D-strip, though containing two spheres instead of one, as depicted schematically in Fig. 3.1 (lower pane). Toward the end of the Chapter, we generalize this configuration to the realistic 2D-tray (upper pane).

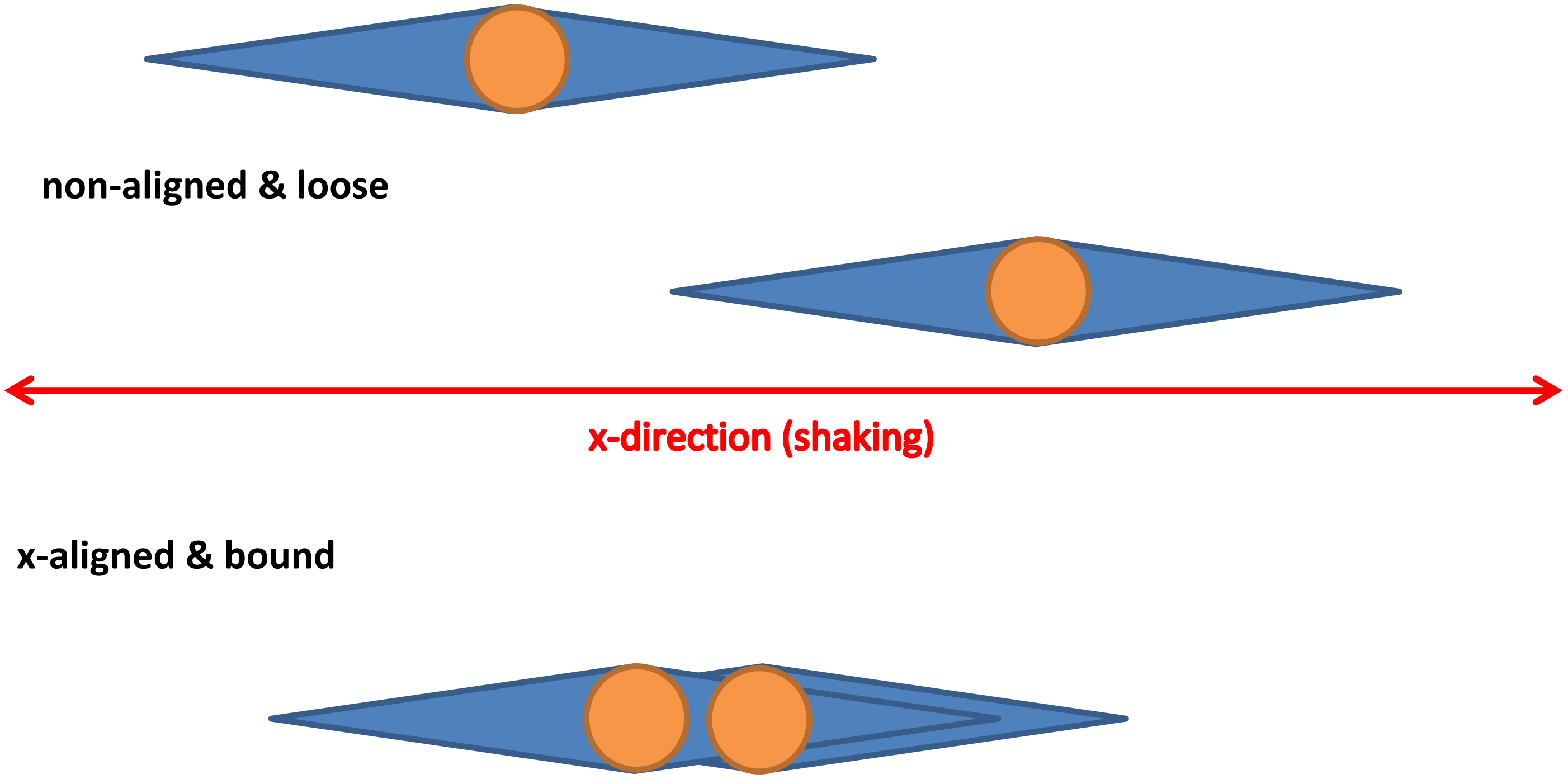

***Figure 1.1: Seed entropic gain upon horizontal alignment of the heavy spheres (orange).*** *The seeds perform a forced motion inside the blue regions, which lowers their entropy. Consequently, reducing the blue surface corresponds to increasing the entropy of the bath of seeds.*

When the sphere-seed difference velocity points rightwards (i.e., when the sphere moves faster rightwards or slower leftwards than the seeds), the seed concentration rises above average in the rightward cone, while it plunges below average in the leftward cone, and *vice versa*. Non-uniform concentrations imply a reduction in entropy of the seed bath, no matter whether those regions have a higher or lower-than-average concentration. For example, zero horizontal-velocity entropy results when all seeds are close-packed at one side of the tray. Consequently, the more uniform the horizontal seed velocity, the higher the horizontal seed velocity entropy. For a better readability of the text, we will henceforth omit the adjective "horizontal", as that is the more important direction in the 2D-plane.

As long as the two spheres have non-overlapping imaginary rhomboids, only long-range Casimir-Polder forces play a role, which are too small, and not exclusively of attractive nature, to explain the experimental observations [24]. As soon as the rhomboids of two different spheres present a tiny overlap, the seeds' entropy increases. This is because in the overlap region, high and low densities superpose, producing an overlap area of equilibrium seed concentration. Fig. 1.1 illustrates the extremal stages of no-overlap and maximum-overlap. The seeds' entropy reduction due to a single sphere is (see Chapter II, Eq. 1.7):

[1.1] $$\frac{\Delta S_{Rr}^{[1D]}(u_0,\omega_h t,N_r)}{S_{r-stat}^{[1D]}(N_r)} = [1+\Delta V_{Rr}^{q}(u_0,\omega_h t,N_r)]^{-1}$$

with $\Delta V_{Rr}(u_0,\omega_h t,N_r)$ the dimensionless difference velocity (see Chapter II, Eq. 1.6), and $N_r$ the number of seeds in the pseudo-2D box. The latter is rewritable in terms of seed concentration $C_r$ and seed nearest-neighbor distance, see Eq. 7.1 of Chapter I). The superscript "1D" (with $W^{[1D]} = 2R$) denotes the pseudo-2D configuration (with $W^{[2D]} >> 2R$). We added the power $q > 0$ as a possible fit parameter, though reality is doubtlessly more complex than this: e.g., there could be an ulterior dependence on concentration. Eq. 1.1, with $q$ set to unity, is the starting point for the formulation of the entropy change in the case of two spheres, which we shall indicate with the suffix "RRr". As long as the regions of sphere-induced variation of seed-concentration do not overlap, the two-sphere differential entropy simply doubles. However, as soon as two rhombic surfaces overlap, as in the pseudo-2D case, the distance-dependent difference entropy of two adjacent spheres in a seed bath reads

[1.2] $$\frac{\Delta S_{RRr}^{[1D]}(L_{RR})}{S_{r-stat}^{[1D]}(N_r)} = \frac{2-(1-\frac{L_{RR}}{L_{\mathrm{Rhomb}}})^2}{1+\Delta V(L_{RR})}\theta(L_{\mathrm{Rhomb}}-L_{RR})\theta(L_{RR}-2R)$$

with $L_{RR}$ the heart-to-heart distance between two horizontally aligned spheres, $L_{\mathrm{Rhomb}}$ the rhombic length (twice the cone length, which is defined as the distance from sphere center to cone extreme). The second Heaviside function appearing in Eq. 1.2 grants that the spheres never overlap in space. The two-sphere generalization of the single-sphere, dimensionless sphere-seed velocity difference $\Delta V_{Rr}(u_0, \omega_h t, N_r)$, given in Chapter II (Eq. 3.5), is

[1.3] $$\Delta V(L_{RR}) \equiv 2\Delta V_{Rr} - \beta(L_{RR}-2R)^2$$

which satisfies the three basic conditions,

[1.4] $$\left\{\begin{array}{l} \Delta V(2R) = 2\Delta V_{Rr} \\ \left.\frac{\partial}{\partial L_{RR}}\Delta V(L_{RR})\right|_{2R} = 0 \\ \Delta V(L_{\mathrm{Rhomb}}) = \Delta V_{Rr} \end{array}\right\}$$

provided that

[1.5] $$\Delta V(L_{\mathrm{Rhomb}}) = 2\Delta V_{Rr} - \beta(L_{\mathrm{Rhomb}}-2R)^2 \quad \Rightarrow \quad \beta \equiv \frac{\Delta V_{Rr}}{(L_{\mathrm{Rhomb}}-2R)^2}$$

Consequently, the first two derivatives of Eq. 1.3 are

[1.6] $$\left\{\begin{array}{l} \frac{\partial}{\partial L_{RR}}\Delta V(L_{RR}) = -2\beta(L_{RR}-2R) \\ \frac{\partial^2}{\partial L_{RR}^2}\Delta V(L_{RR}) = -2\beta \end{array}\right\}$$

The second derivative of the entropy is

[1.8] $$\left\{\begin{array}{l} f(L_{RR}) \equiv \frac{\partial^2}{\partial L_{RR}^2} \frac{\Delta S_{RRr}^{[1D]}(L_{RR})}{S_{r-stat}^{[1D]}(N_r)} = -\frac{2}{(1+\Delta V)L_{\mathrm{Rhomb}}^2} \times \\ \times \left[ 2\frac{L_{\mathrm{Rhomb}} - L_{RR}}{1+\Delta V} \Delta V' + 1 + [\tfrac{1}{2}(L_{\mathrm{Rhomb}}^2 - L_{RR}^2) + L_{\mathrm{Rhomb}} L_{RR}][\frac{\Delta V''}{1+\Delta V} - 2(\frac{\Delta V'}{1+\Delta V})^2 \right] \end{array}\right\}$$

Before diving into thermodynamics, we first deduce the average contact time from experiment [24].

## 3.2. Measurement of the Average Contact Time

The stability of the bi-sphere bond depends on the depth of their entropy well as compared to the average horizontal kinetic energy of the spheres. The average time the two spheres stay bonded follows directly from Lozano's measured $L_{CCD}(\lambda)$ -curves in Fig. 5 [24], with the definitions.

[2.1] $$\left\{\begin{array}{l} \lambda \equiv \ln \tau \\ L_{CCD} \equiv \ln F_{CCD} \end{array}\right\}$$

The subscript CCD stands for "Complementary Cumulative Distribution".

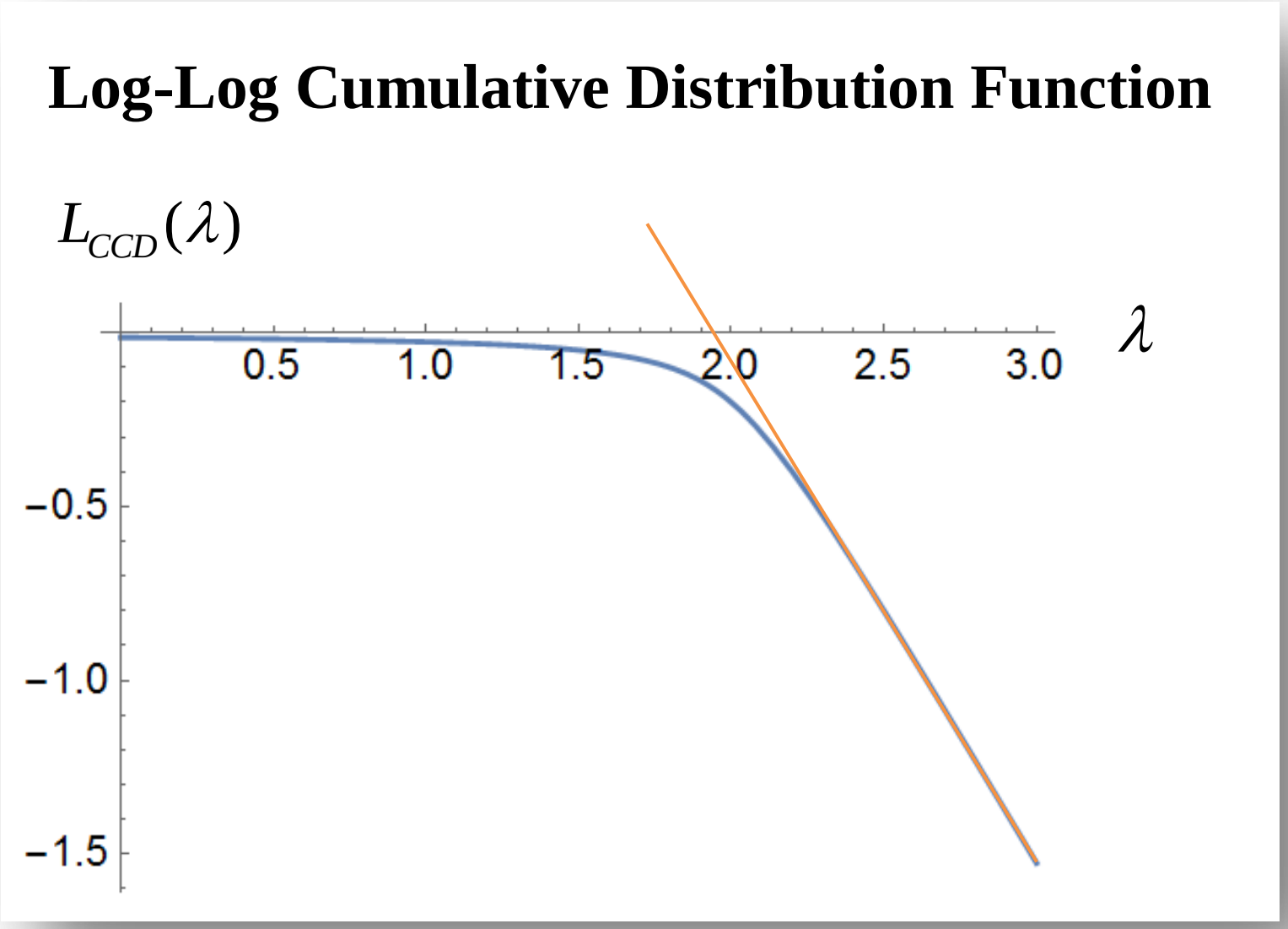

***Fig. 2.1: The hyperbolic*** $L_{CCD}(\lambda)$ ***fitted to Lozano's blue curve of Fig. 3 of Supplementary Information, Ref. 24.*** *The crossing time of the two hyperbolic axes follows from [24] as* $\tau_{hyp} = 10^{0.8} = 6.8 \quad \Rightarrow \quad \lambda_{hyp} = \ln 6.8 = 1.92$. *Note that both axes represent natural-base logarithms, in contrast to Ref. 24, which uses the decimal base.*

The purple curve in Fig. 3 of the Lozano's SI [24] shows $L_{CCD}(\lambda)$. It has the shape of the hyperbola

[2.2] $$[\frac{h}{L_{CCD}(\lambda)}]^2 - (\alpha - 1)\frac{\lambda - \lambda_{hyp}}{L_{CCD}(\lambda)} = 1$$

Fig. 2.1 illustrates a hyperbolic fit with $h = -0.2$, and $\alpha = 2.37$. Evaluation of Eq. 4.8 at $\lambda = \lambda_{hyp}$ yields the definition of *h*

[2.3] $$h = \pm L_{CCD}(\lambda_{hyp})$$

The the skew hyperbolic symmetry axis follows from the limit $h \downarrow 0$ of Eq. 2.2

[2.4] $$H_{skew}(\lambda) = \lim_{h \to 0} L_{CCD}(\lambda) = -(\alpha - 1)(\lambda - \lambda_{hyp})$$

That means, it has a logarithmic slope of $-(\alpha-1)$ and intersects the other hyperbolic axis at $\lambda_{hyp}$:

$$[2.5] \quad \begin{Bmatrix} L_{CCD}^2(\lambda) - 2q(\lambda)L_{CCD}(\lambda) = h^2 \\ [L_{CCD}(\lambda) - q(\lambda)]^2 = h^2 + q^2(\lambda) \\ L_{CCD}(\lambda) = q(\lambda) - u(\lambda) \end{Bmatrix}$$

with the definitions

$$[2.6] \quad \begin{Bmatrix} q(\lambda) \equiv -\tfrac{1}{2}(\alpha-1)(\lambda-\lambda_{hyp}) \\ u(\lambda) \equiv \sqrt{h^2 + q^2(\lambda)} \end{Bmatrix}$$

The positive sign in Eq. 2.6 for $u(\lambda)$ selects the negative hyperbolic branch; the second hyperbolic axis follows in the limit $h \downarrow 0$: it coincides with the abscissa. From Eq. 2.5, it follows that

$$[2.7] \quad F_{CCD}(\lambda) = e^{q(\lambda)-u(\lambda)}$$

Fig. 2.2 illustrates the function 2.7:

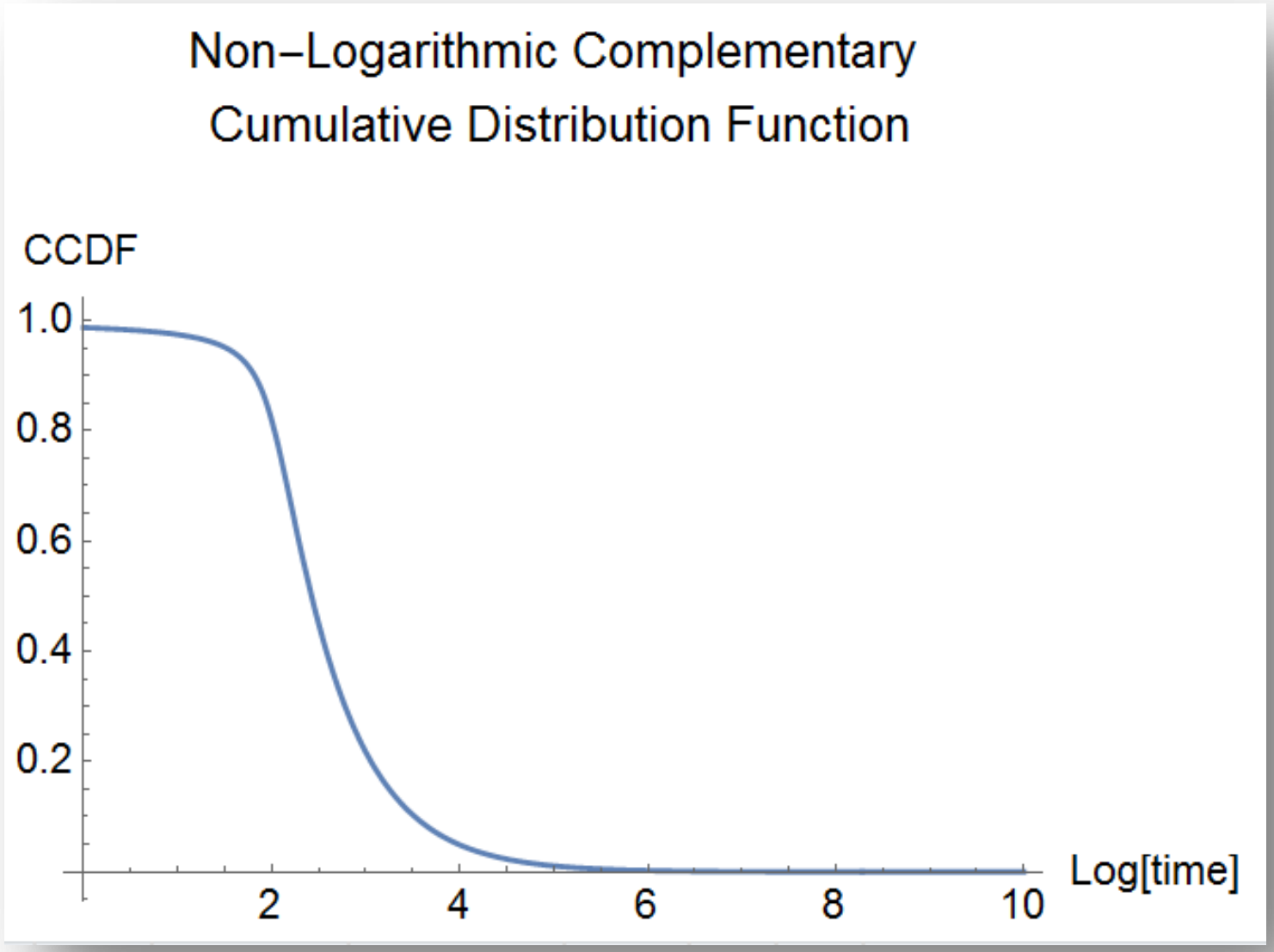

***Figure 2.2: Log-Lin representation of the complementary cumulative distribution function*** $F_{CCD}(\lambda)$**.** *Both fit parameters,* $\lambda_{hyp} = 1.92$, *and* $\alpha = 2.37$, *are unrecognizable.*

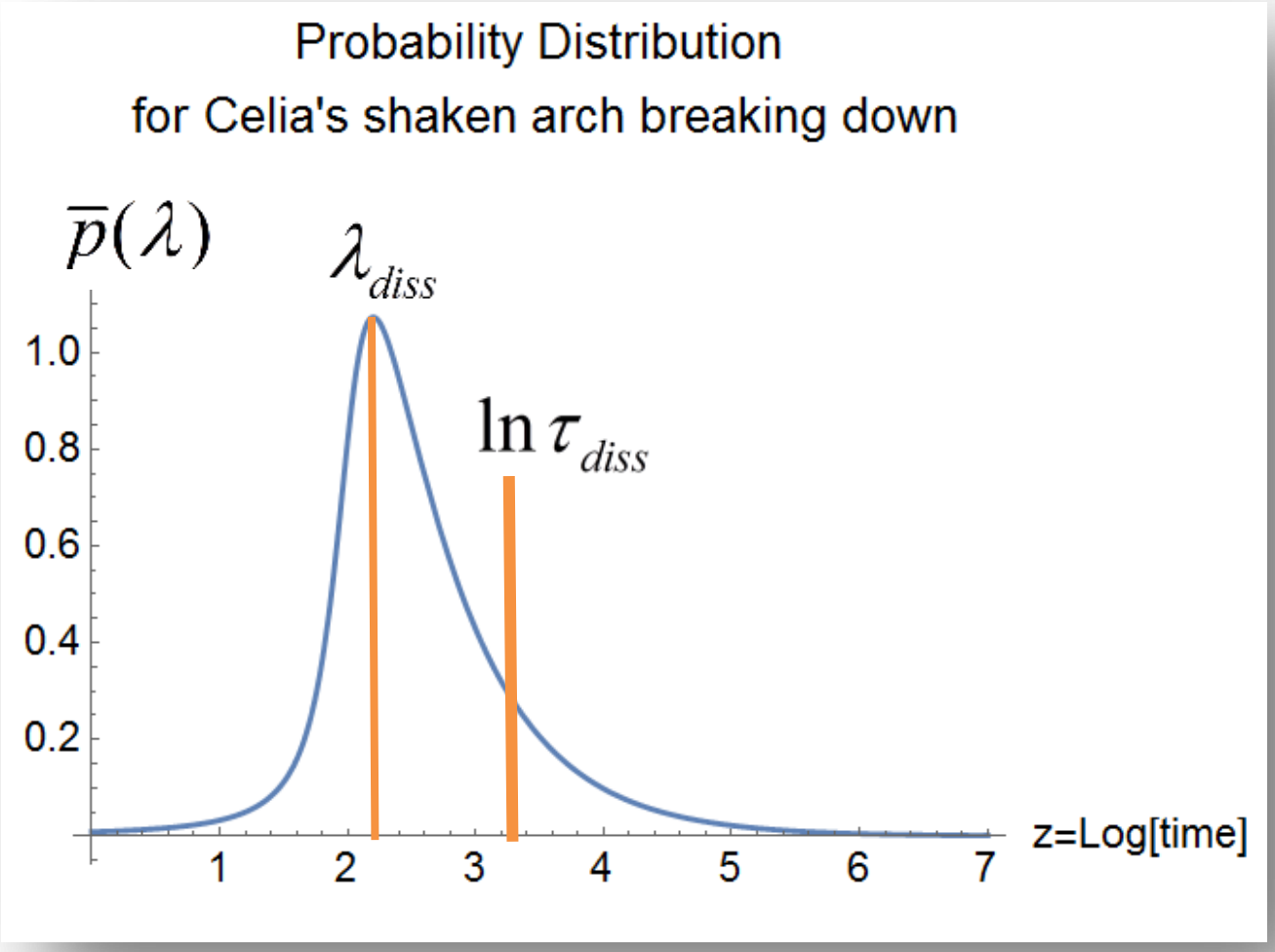

***Figure 2.3: Log-Lin representation of the probability distribution,*** $\bar{p}(\lambda)$***.*** *The thick orange line represents the average value (26 =* $e^{3.3}$*) of the linear distribution. It exceeds by 50% the average of the logarithmic maximum, located around* $e^{2.4}$*, and represented by the thin orange line.*

As $F_{CCD}(\lambda)$ is a cumulative distribution, its derivative yields the probability distribution

$$[2.8] \qquad \bar{p}(\lambda) \equiv -\frac{\partial}{\partial\lambda}F_{CCD}(\lambda) = -\frac{\partial e^{-q-u}}{\partial(q+u)}\frac{\partial(q+u)}{\partial q}\frac{\partial q}{\partial\lambda} = \frac{1}{2}(\alpha-1)e^{-q-u}(1+\frac{q}{u})$$

with omitted explicit $\lambda$-dependences of the two functions $q$ and $u$. The probability distribution $\bar{p}(\lambda)$ is normalized to unity by definition, because its integral, $F_{CCD}(\lambda)$, has unit value at $\lambda = 0$:

$$[2.9] \qquad \int_0^\infty d\lambda\, \bar{p}(\lambda) = F_{CCD}(0) = 1$$

The probability distribution $p(\tau)$, as a function of dimensionless time $\tau$, satisfies

$$[2.10] \qquad \left\{ \begin{array}{l} d\lambda\, \bar{p}(\lambda) = d\tau\, p(\tau) \\ p(\tau) = \bar{p}(\lambda)\dfrac{d\lambda}{d\tau} = \bar{p}(\lambda)\dfrac{d\ln\tau}{d\tau} \\ \bar{p}(\ln\tau) = \tau\, p(\tau) \end{array} \right\}$$

The average dissociation time $\tau_{diss}$ is the first moment of $p(\tau)$:

$$[2.11] \quad \tau_{diss} = \int_0^\infty \tau d\tau p(\tau) = \int_0^\infty d\tau \overline{p}(\ln\tau) = \frac{1}{2}(\alpha - 1)\int_0^\infty d\tau\, e^{-q-u}(1+\frac{q}{u})$$

The numerical integral yields $t_{diss} = 26s$.

## 3.3. Thermodynamics on a Cylinder Surface

The Gibbs free energy potential of thermodynamics is a function of temperature *T*, pressure *P*, and particle numbers:

$$[3.1] \quad dG(T,P,\{N_j\}) = -SdT + VdP + \mu_j \sum_{j}^{species} dN_j$$

With $\mu_j$ and $N_j$ the particles' chemical potential and number, respectively, *S* the system's entropy, and *V* its volume. Applied to our 2D-case of constant particle numbers, it reduces to

$$[3.2] \quad d\Delta G(T,P) = -\Delta S_{rRR}^{[2D]} dT_x + WLd\Delta P_r^{[2D]}$$

The reader should keep in mind that the temperature is due only to kinetic and angular energy of the particles, and pressure only to their kinetic energy, a rather crude approximation stemming from the absence of internal degrees of freedom. Consequently, all heating energy goes exclusively into the driver's traction band. We use a capital *P* for pressure in order to distinguish it from linear momentum (*p*), and find that, to first order, the pressure is independent of the distance between the spheres.

$$[3.3] \quad WLP_r^{[2D]}(\omega_h t) = 2N_r m_r [v_{rx}^{[rel]}(\omega_h t)]^2 \quad \Rightarrow \quad P_r^{[2D]}(\omega_h t) \equiv 2m_r [\frac{v_{rx}^{[rel]}(\omega_h t)}{d_r}]^2$$

with $d_r^2$ the average available 2D-area for a single seed. In the first equality, the factor 2 is due to the number of transverse borders containing the particle pressure. Since the pressure does not depend on the distance between the spheres, the attractive force between two spheres follows by deriving the Gibbs energy to $L_{RR}$:

$$[3.4] \quad \left\{ \begin{array}{l} F_{attr}^{[1D]}(L_{RR}) = -T_x S_{r-stat}^{[1D]} \dfrac{\partial}{\partial L_{RR}} \dfrac{\Delta S_{RRr}^{[1D]}(L_{RR})}{S_{r-stat}^{[1D]}} = \\ = \dfrac{2}{L_{\text{Rhomb}}} \dfrac{1 - \dfrac{L_{RR}}{L_{\text{Rhomb}}}}{1+\Delta V} - \dfrac{2-(1-\dfrac{L_{RR}}{L_{\text{Rhomb}}})^2}{(1+\Delta V)^2} \Delta V' \end{array} \right\}$$

A similar argument holds for the Gibbs eigenfrequency levels. The top eigenfrequency follows from Eq. 1.8 as

[3.5] $$\begin{Bmatrix} \omega_{top}^2 \equiv -\left.\dfrac{\partial^2 \Delta G(L_R)}{\mu_{Rr}\partial L_{RR}^2}\right|_{L_{\text{Rhomb}}} = \dfrac{T_x S_{r-stat}^{[1D]}(N_r)}{\mu_{Rr}} f(L_{\text{Rhomb}}) = \\ = 2\dfrac{T_x S_{r-stat}^{[1D]}}{\mu_{Rr}(1+\Delta V)L_{\text{Rhomb}}^2}\left( L_{\text{Rhomb}} L_{RR}[2(\dfrac{\Delta V'}{1+\Delta V})^2 - \dfrac{\Delta V''}{1+\Delta V}] - 1 \right) \end{Bmatrix}$$

with $\mu_{Rr} \equiv \dfrac{m_R m_r}{m_R + m_r}$ the reduced mass. For the symbols not discussed in this Section, we refer the reader to Chapter II, first and fourth Sections. The average valley eigenfrequency of the entropic chapter of the Gibbs free energy landscape depends on the sphere density. In the pseudo-1D case, the spheres are circles, and all circles align inside circular 1D-strips with a circumference of $L$ and a height of $2R$. The sphere concentration relates to the number of spheres as

[3.6] $$LC_R^{[1D]} = 2RN_R^{[1D]}$$

Consequently, the average distance between two horizontally aligned spheres (on a circular strip) is

[3.7] $$\overline{L_{RR}^{[1D]}} = \frac{L}{N_R^{[1D]}}$$

Attraction occurs as soon as this value falls below the rhombic long axis, $L_{\text{Rhomb}}$, because at that moment the valley curvature becomes negative instead of flat. Since on the 1D-strip that is necessarily the case

[3.8] $$\begin{Bmatrix} \omega_{valley}^2 \equiv \left.\dfrac{\partial^2 \Delta G(L_R)}{\mu_{Rr}\partial L_{RR}^2}\right|_{2R} = \dfrac{T_x S_{r-stat}^{[1D]}(N_r)}{\mu_{Rr}} f(2R) = \dfrac{2}{(1+\Delta V)L_{\text{Rhomb}}^2} \times \\ \times \left[ 2\dfrac{L_{\text{Rhomb}} - 2R}{1+\Delta V}\Delta V' + 1 + [\tfrac{1}{2}(L_{\text{Rhomb}}^2 - 4R^2) + 2L_{\text{Rhomb}}R][\dfrac{\Delta V''}{1+\Delta V} - 2(\dfrac{\Delta V'}{1+\Delta V})^2 \right]_{2R} \end{Bmatrix}$$

From Section 1 the derivatives of the dimensionless velocity differences are known

[3.9] $$\begin{Bmatrix} \Delta V' = -2\beta(L_{RR} - 2R) \\ \Delta V'' = -2\beta \end{Bmatrix} \quad \& \quad \beta \equiv \frac{\Delta V_{Rr}}{(L_{\text{Rhomb}} - 2R)^2}$$

## 3.4. Generalization to a 2D-tray

The generalization to two dimensions essentially requires calculating the rhombic overlap as a function of both lateral and longitudinal distances between two nearest neighbor spheres. In order not to complicate the equations unnecessarily, we stick to rectangles instead of rhombs:

[4.1] $$O_{rect}^{[2D]}(N_{R-tot}, L_{RR}, W_{RR}) \equiv \alpha_0 N_{R-tot}(L_{\mathrm{rect}} - L_{RR})(2R - W_{RR})\theta(L_{\mathrm{rect}} - L_{RR})\theta(2R - W_{RR})$$

The symbol *O* acquires the meaning of an average overlap fraction, upon choosing its normalization such that

[4.2] $$\int_0^L dL_{RR} \int_0^W dW_{RR}\, O_{rect}^{[2D]}(\frac{LW}{4R^2}, L_{RR}, W_{RR}) \equiv 1$$

This equation means that, when the whole available tray surface is filled with only phosphor-bronze spheres, in a checker-board configuration, the overlap fraction becomes unity. This condition requires

[4.3] $$\alpha_0 \frac{LW}{4R^2} \int_0^L dL_{RR} \int_0^W dW_{RR}\, (L_{\mathrm{rect}} - L_{RR})(2R - W_{RR}) = 1$$

Thus setting the normalization constant to

[4.4] $$\alpha_0 \equiv \frac{4}{L_{rect}^2 LW} (\frac{L}{L_{\mathrm{rect}}} - 1)^{-2} (\frac{W}{2R} - 1)^{-2}$$

The most probable initial sphere distribution is such that both longitudinal and transversal distance distributions are Gaussian:

[4.5] $$\left\{ \begin{array}{l} \rho_{long}^{[2D]}(L_{RR}) \equiv \dfrac{2}{d_{RR}\sqrt{\pi}} e^{-(\frac{L_{RR}}{d_{RR}})^2} \\ \rho_{trans}^{[2D]}(W_{RR}) \equiv \dfrac{2}{d_{RR}\sqrt{\pi}} e^{-(\frac{W_{RR}}{d_{RR}})^2} \end{array} \right\}$$

such that for the limit that *L* and *W* (the integration maxima) go to infinity, the integrals go to unity. At low sphere concentration, the average distance between two spheres is equal to:

[4.6] $$d_{RR}^2(N_{R-tot}) \equiv \frac{LW}{N_{R-tot}} \quad \Rightarrow \quad \rho_{square}^{[2D]}(L_{RR},W_{RR}) \equiv \rho_{long}^{[2D]}(L_{RR})\rho_{trans}^{[2D]}(W_{RR}) = \frac{4N_{R-tot}}{\pi LW} e^{-N_{R-tot}\frac{L_{RR}^2+W_{RR}^2}{LW}}$$

For all possible sphere concentrations, $N_{R-tot} \equiv \sum_{k=1}^{\infty} kN_{kR}$, with $N_{RR} = N_{2R}$ the number of bi-spheres. For small enough sphere concentrations (such that tri-spheres do not exist) as

[4.7] $$\begin{cases} N_{RR}(N_{R-tot},L,W) = LW\int_0^L dL_{RR} \int_0^W dW_{RR}\, \rho_{square}^{[2D]}(L_{RR},W_{RR}) O_{rect}^{[2D]}(N_{R-tot},L_{RR},W_{RR}) = \\ = 32N_{R-tot}\frac{R}{L_{\mathrm{rect}}}(\frac{L}{L_{\mathrm{rect}}}-1)^{-2}(\frac{W}{2R}-1)^{-2}\mathrm{Erf}\,(L_{\mathrm{rect}}\sqrt{\frac{N_{R-tot}}{LW}})\mathrm{Erf}\,(2R\sqrt{\frac{N_{R-tot}}{LW}}) \end{cases}$$

given the standard integral

[4.8] $$\int_{-y}^{y} dx\, e^{-\beta x^2}(y-x) = y\int_{-y}^{y} dx\, e^{-\beta x^2} = y\sqrt{\frac{\pi}{\beta}}\mathrm{Erf}\,(y\sqrt{\beta})$$

Of course, the number of bi-spheres can be a fractional number. In Lozano's experiment it would be equal to the ratio of bound time and the sum of bound and free times:

[4.9] $$N_{RR}(2,L,W) = \frac{\tau_{bound}}{\tau_{bound} + \tau_{free}}$$

Hence, equating Eqs. 4.9 and 4.7 immediately yields the rhombic long axis.

## 3.5. Retrieval of physical parameters using Kramers' rate equation

The exact meaning of the rhombic (or rectangular) long axis is not yet clear. As has been said when this quantity was first introduced, it is directly connected to the longitudinal extent of influence that the sphere has on the motion of the seeds. Evidently, there is a longitudinal density gradient along the rectangular or rhombic axis, too. A crude estimate of the average energy barrier results upon considering the entropy independent of temperature:

[5.1] $$\left\{\begin{array}{l} \overline{G_b}(\omega_h t) \equiv \Delta G(\omega_h t, L_{\text{Rhomb}}, T_{rx-RR}) - \Delta G(\omega_h t, 2R, T_{rx-RR}) = \\ = -\int_0^{T_{rx-RR}} dT\, [\Delta S_{rRR}^{[2D]}(\omega_h t, L_{\text{Rhomb}}) - \Delta S_{rRR}^{[2D]}(\omega_h t, 2R)] = \\ = [\Delta S_{rRR}^{[2D]}(\omega_h t, 2R) - \Delta S_{rRR}^{[2D]}(\omega_h t, L_{\text{Rhomb}})] T_{rx-RR} \end{array}\right\}$$

with

[5.2] $$\tfrac{1}{2} k_B T_{rx-RR}(t, u_0) = \tfrac{1}{2} N_r m_r [v_r^{[rel]}(t, u_0)]^2 + m_R [v_R^{[rel]}(t, u_0)]^2$$

The factor one-half of the LHS indicates a single degree of freedom (kinetic energy along the longitudinal direction), and the absence of that factor in the second term of the RHS is due to the fact that two spheres are summed ( $N_{R-tot} = 2$ ). We now use the Eigenfrequency $\omega_{bound}$ for the valley of two bound spheres, and $\eta$ the effective seed bath viscosity. In case $\eta \gg \omega_{bound}$ , Kramers' general transition rate equation [29] simplifies to

[5.3] $$-\frac{\eta}{2} + \sqrt{\left(\frac{\eta}{2}\right)^2 + \omega_{bound}^2} \approx \frac{\omega_{bound}^2}{\eta} \quad \& \quad \left\{\begin{array}{l} r_{diss} \equiv \dot{N}_{bound \to free} \approx N_{bound} \dfrac{\omega_{bound}\omega_{top}}{2\pi\eta} e^{-\frac{G(L_{rect}) - G(2R)}{k_B T}} \\ \dot{N}_{free \to bound} \approx N_{free} \dfrac{\omega_{free}\omega_{top}}{2\pi\eta} e^{-\frac{GU_{free}}{k_B T}} \end{array}\right\}$$

The rate $r_{diss} \equiv t_{diss}^{-1}$ symbolizes the average dissociation rate of sphere pairs. In thermodynamic equilibrium the two rates in Eq. 5.3 are equal per definition

[5.4] $$\frac{N_{RR}}{N_{1R}} \equiv 2 \frac{N_{bound-pair}}{N_{free-pair}} = 2 \frac{\omega_{free} \exp[\frac{\Delta G_{free}}{k_B T}]}{\omega_{bound} \exp[\frac{\Delta G_{bound}}{k_B T}]}$$

with the definitions

[5.5] $\Delta G_{bound} \equiv G(L_{rect}) - G(2R)$ & $\Delta G_{free} \equiv G_{free}(L_{\mathrm{rhomb}}) - G_{free}(L_{mx}) \equiv -G_{free}(L_{mx})$

Here, $L_{mx}$ stands for half the distance between two neighboring ($2R < L_{RR} < L_{\mathrm{rhomb}}$) spheres. For two spheres, one has

[5.6] $$\begin{cases} \omega_{bound} \gg \omega_{free} \approx 0 \\ \Delta G_{bound} \gg \Delta G_{free} \approx 0 \end{cases}$$

granting that both factors in Eq. 5.5 cause the time-averaged number of free sphere pairs to exceed by far that of bound pairs. Finally, it should be noted that all quantities in this Section vary harmonically with the driver's oscillation frequency. Fig. 5.1 illustrates the Gibbs free energy landscape for three spheres at different distances: the third is at outside its neighbor's rhombic distance of influence en therefore is subject to only Brownian motion induced by the temperature of the harmonically swept bath of seeds. The Brownian motion of the seeds and the spheres themselves determine the odds that a given approach of the first two spheres eventually leads to a bi-sphere cluster or not.

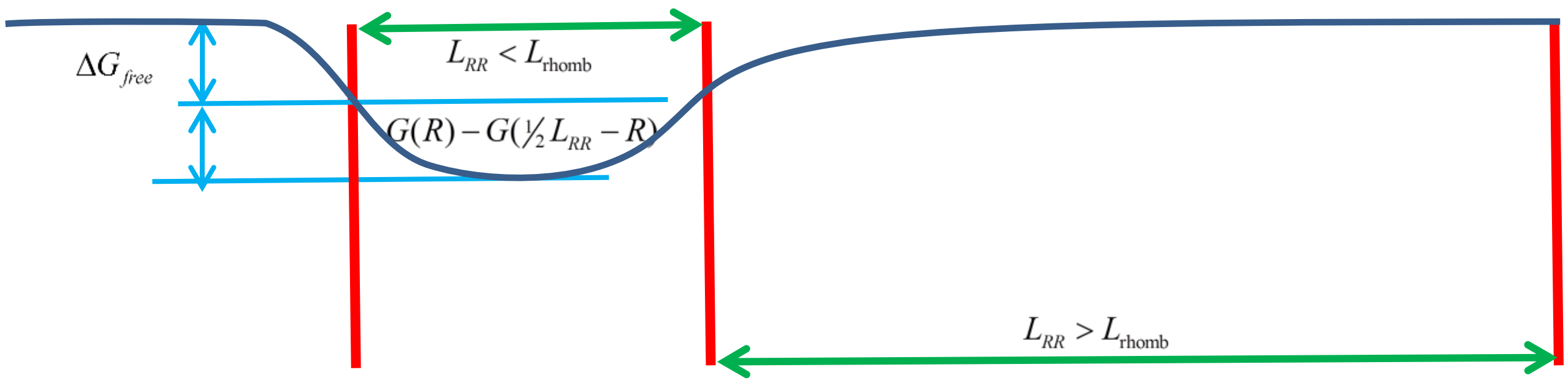

***Figure 5.1: Gibbs free energy landscape (dark blue) for three transversally aligned spheres (red vertical position lines).*** *The Gibbs free energy vanishes at the highest Gibbs energy level. The difference energies are positive by definition. Whenever* $|L_{RR}| > L_{\text{rhomb}}$*, there exists no entropic attractive force between adjacent spheres. In the opposite case, however, the attractive force increases rapidly with decreasing distance, while reducing* $\Delta G_{bound}$ *, and increasing* $\Delta G_{free}$*.*

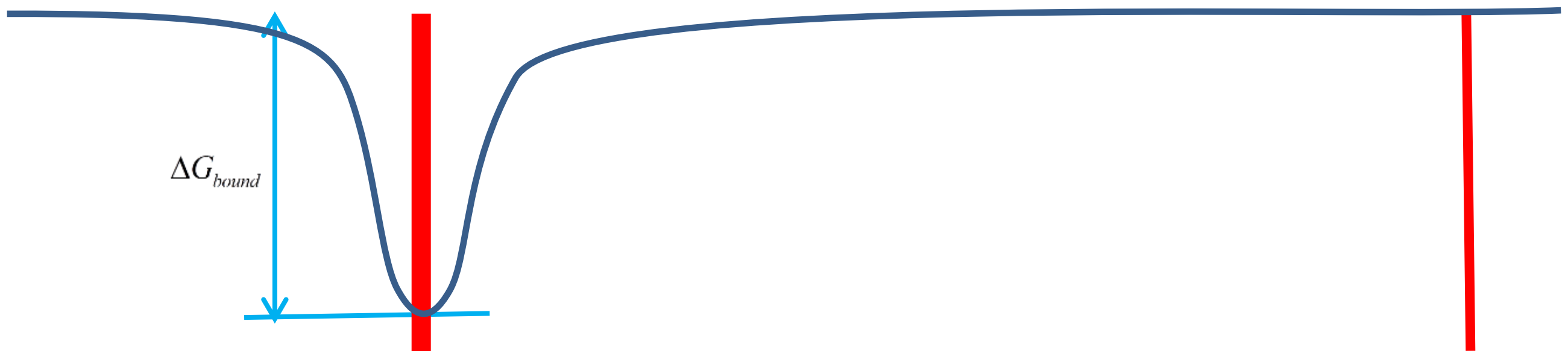

***Figure 5.2: Gibbs free energy landscape (dark blue) for three transversally aligned spheres (red vertical position lines).*** *Whenever the entropic attraction wins against the Brownian motion, the first to spheres unite to form ma bi-sphere cluster. Else, the second sphere might eventually have united with the third one. A bi-sphere survives until Brownian motion tears the two spheres far enough apart.*

## 3.6. Conclusions

In Chapter I of this paper, we present a theoretical explanation for the experimentally measured longitudinal diffusion of the spheres [24] as a function of seed concentration. The rather good agreement confirms our initial Ansatz that the sphere's rotational and longitudinal velocity have at all times the same phase. We do not require strict energy partitioning between the sphere's kinetic and rotational energy, but a ratio between the two, which is independent of time. Whatever ratio best fits the experimental results, the mere existence of a fixed ratio has the enormous advantage that the problem analytically treatable using nineteenth century mathematics.

Chapter II takes advantage of the mathematical simplification of the equal-phase Ansatz to propose an equation of the seed bath's velocity and resulting entropy, as well as for the sphere in a seed bath. All physical quantities can now be analytically calculated with little numerical effort, allowing for 3D-contour plots of those quantities, as a function of those physical parameters that the experimentalist is free to choose.

Chapter III details the relations between dissociation and binding rates, seed bath viscosity, the Gibbs' energy landscape, along with its top and valley eigenfrequency spacings.

## Appendix 1: Specific parameters of the experiments performed in Ref. 24.

Length and width of the tray: $\left\{ \begin{array}{l} L = 180\,mm \\ W = 90\,mm \end{array} \right\}$

Driver frequency (periods per second) and velocity amplitude: $\left\{ \begin{array}{l} \nu_h \equiv 12\,Hz \\ u_0 \equiv x_{tray}^{[ampl]} \omega_h = 130\,mms^{-1} \end{array} \right\}$

Average radius, mass density and mass of the poppy seed: $\left\{ \begin{array}{l} r \equiv 0.53\,mm \\ \rho_r = 0.2\,gcm^{-3} \\ m_r = 1.2 \times 10^{-7}\,kg \end{array} \right\}$

Likewise for the phosphor-bronze spheres: $\left\{ \begin{array}{l} R \equiv 0.75\,mm \\ \rho_r = 8.8\,gcm^{-3} \\ m_R = 1.6 \times 10^{-5}\,kg \end{array} \right\}$

# Appendix 2: Sphere's Amplitude and Phase Delay in Seed Bath

The normalization of the Kernel (Eq. II.2.13) is

$$[A.1] \quad \begin{cases} M \equiv \int_0^{\omega_h t_{ff}} dz\, K(z, C_r, \zeta_\mu) = -\frac{2}{3C_r} Q_{ff} \log Q_{ff} - \frac{(\omega_h t_{ff})^2}{2\zeta_\mu} \\ Q_{ff} \equiv 1 - \frac{3C_r \omega_h t_{ff}}{2} \quad \& \quad 0 < Q_{ff} < 1 \end{cases}$$

and the non-normalized velocity is

$$[A.2] \quad \begin{cases} M \frac{v_{Rxh}^{[abs]}(\omega_h t, C_r, \zeta_\mu)}{u_0} = \mathrm{Re}\, e^{i\omega_h t} \int_0^{\omega_h t_{ff}} dz\, e^{-iz} K(z, C_r, \zeta_\mu) = \\ = \mathrm{Re}\, e^{i\omega_h t} \left[ \frac{1 - e^{-i\omega_h t_{ff}} \left(1 + i\omega_h t_{ff}\right)}{\zeta_\mu} + i\left(\cos\omega_h t_{ff} - 1\right) + \sin\omega_h t_{ff} + is + ie^{-i\omega_h t_{ff}} \log Q_{ff} \right] \end{cases}$$

with, in terms of exponential integrals EI,

$$[A.3] \quad s \equiv e^{-\frac{2i}{3C_r}} \left( \mathrm{EI}[\frac{2i}{3C_r}] - \mathrm{EI}[\frac{2iQ_{ff}}{3C_r}] \right)$$

The amplitude and phase follow from regrouping the above terms:

$$[A.4] \quad \begin{cases} V_h \equiv \frac{A_m}{M} \\ A_m \equiv \sqrt{A_c^2 + A_s^2} \\ \phi_h \equiv \arccos\frac{A_c}{A_m} \mathrm{sgn}[\arcsin\frac{A_s}{A_m}] \end{cases}$$

with

$$[A.5] \quad \begin{cases} A_s \equiv 1 - \cos\omega_h t_{ff} + \frac{\omega_h t_{ff} \cos\omega_h t_{ff} - \sin\omega_h t_{ff}}{\zeta_\mu} - \cos\omega_h t_{ff} \log Q_{ff} - s' - \sin\omega_h t_{ff} \\ A_c \equiv \sin\omega_h t_{ff} + \frac{1 - \cos\omega_h t_{ff} - \omega_h t_{ff} \sin\omega_h t_{ff}}{\zeta_\mu} + \sin\omega_h t_{ff} \log Q_{ff} - s'' - \cos\omega_h t_{ff} \end{cases}$$

# Acknowledgements

The author gratefully acknowledges the department head, Iker Zuriguel, for introducing me into his interesting field of research, for valuable discussions, and for granting me access to his laboratories.